\documentclass[10pt,letterpaper]{article}
\usepackage[top=0.85in,left=2.75in,footskip=0.75in,marginparwidth=2in]{geometry}

\usepackage[utf8]{inputenc}

\usepackage{cite}
\usepackage{url}

\usepackage{nameref,hyperref}

\usepackage[right]{lineno}

\usepackage{microtype}
\DisableLigatures[f]{encoding = *, family = * }

\raggedright
\usepackage{changepage}

\usepackage[aboveskip=1pt,labelfont=bf,labelsep=period,singlelinecheck=off]{caption}

\makeatletter
\renewcommand{\@biblabel}[1]{\quad#1.}
\makeatother

\usepackage{lastpage,fancyhdr,graphicx}
\usepackage{epstopdf}
\usepackage{algorithmic}
\usepackage{algorithm}

\usepackage{amsmath,amsthm}
\usepackage{epstopdf}
\fancyheadoffset[L]{2.25in}
\fancyfootoffset[L]{2.25in}

\usepackage{color}

\definecolor{Gray}{gray}{.25}

\usepackage{graphicx}

\usepackage{sidecap}

\usepackage{wrapfig}
\usepackage[pscoord]{eso-pic}
\usepackage[fulladjust]{marginnote}
\reversemarginpar

\begin{document}
\vspace*{0.35in}

% title goes here:
\begin{flushleft}
{\Large
\textbf\newline{Predicting Parkinson's disease evolution using deep learning.}
}
\newline
% authors go here:
\\
Maria Frasca\textsuperscript{1,*},
Davide La Torre\textsuperscript{2},
Gabriella Pravettoni\textsuperscript{1}
Ilaria Cutica\textsuperscript{1},

\bigskip
\bf{1} Department of Oncology and Hemato-Oncology, University of Milan, Italy
\\
\bf{2} SKEMA Business School, Univerisit\'{e} C\^{o}te d'Azur, Sophia Antipolis, France
\\
\bigskip
* maria.frasca@unimi.it

\end{flushleft}

\section*{Abstract}
Parkinson's disease (PD) is a neurological condition that occurs in nearly 1\% of the world's population.
The disease is manifested by a sharp drop in dopamine production, symptoms are cognitive and behavioural and include a wide range of personality changes, depressive disorders, memory problems, and emotional dysregulation, which can occur as the disease progresses. Early diagnosis and accurate staging of the disease are essential to apply the appropriate therapeutic approaches to slow cognitive and motor decline.

Currently, there is not a single blood test or biomarker available to diagnose Parkinson's disease. 
Magnetic resonance imaging (MRI) has been used for the past three decades to diagnose and distinguish between PD and other neurological conditions. However, in recent years new possibilities have arisen: several AI algorithms have been developed to increase the precision and accuracy of differential diagnosis of PD at an early stage.

To our knowledge, no AI tools have been designed to identify the stage of progression.
This paper aims to fill this gap. Using the "Parkinson's Progression Markers Initiative" dataset, which reports the patient's MRI and an indication of the disease stage, we developed a model to identify the level of progression. 
The images and the associated scores were used for training and assessing different deep-learning models. Our analysis distinguished four distinct disease progression levels based on a standard scale (Hoehn and Yah scale). The final architecture consists of the cascading of a 3DCNN network, adopted to reduce and extract the spatial characteristics of the RMI for efficient training of the successive LSTM layers, aiming at modelling the temporal dependencies among the data. 
 Our results show that the proposed 3DCNN + LSTM model achieves state-of-the-art results by classifying the elements with 91.90\% as macro averaged OVR AUC on four classes

% the * after section prevents numbering
\section*{Introduction}
\label{sec:introduction}
Parkinson's disease (PD) is a neurological condition that is seen more frequently in the elderly. Nearly 1\% of the world population is affected by PD and several affected patients have significant physical and cognitive problems\cite{kouli2018parkinson}.
The disease is manifested by a sharp drop in dopamine production, resulting from the death of the related producing cells present in an area called the substantia nigra present in the midbrain\cite{chu2002age}. The disease most likely spreads throughout the brain via a protein called alpha-synuclein, which is found in the region between the spinal cord and the brain\cite{wakabayashi2020and}.
Symptoms are both cognitive and behavioural, and include a wide range of personality changes, depressive disorders, memory problems, and emotional dysregulation, which can occur as the disease progresses~\cite{poewe2008non}. Also, the symptoms related to mobility worsen as the disease progresses. Early diagnosis is essential to apply the appropriate therapeutic approaches to slow cognitive decline. Currently, a series of tests are needed to diagnose or verify the progression of Parkinson's, as today we are still looking for a biomarker for a quick diagnosis~\cite{frasca2022visualizing,hawkes2010timeline}.

Clinicians identify the disease by analyzing the patient's symptoms, which include slow movement, stiffness, tremors, and problems with balance and coordination. However, these symptoms and the speed at which they progress can vary from case to case~\cite{maiti2017current}.

Magnetic Resonance Imaging (MRI), a non-invasive medical technology that produces images of the internal anatomy of the body, has been used for the past three decades to diagnose and distinguish between PD and other neurological conditions~\cite{heim2017magnetic, kinney2019high}. Furthermore, it has also been used in clinical trials to evaluate the efficacy of therapies~\cite{burciu2017progression}. 

Many clinical studies have tried to detect PD, through different classification methods \cite{saravanan2022systematic,wodzinski2019deep,rajanbabu2022ensemble} and on different types of data, ranging from physiological signals, such as speech \cite{khaskhoussy2023improving} and EEG \cite{bhurane2022diagnosis} to gait \cite{wahid2015classification}, handwriting \cite{rosenblum2013handwriting}, and neuroimaging \cite{zeng2017differentiating} analysis. In particular, neuroimaging analysis has been successfully applied in various sophisticated methods for PD diagnosis\cite{saravanan2022systematic}. A key challenge is to construct quantitative models of pathological and cognitive decline in PD progression using patient data such as neuroimaging and clinical measures for several reasons, including\cite{balestrino2020parkinson}:

\begin{itemize}
      \item To perform an accurate diagnosis. Understanding the progression of PD can help doctors diagnose the disease earlier and more accurately, allowing treatment to be started as soon as possible. This can improve the patient's quality of life and delay the progression of symptoms.
      \item To improve symptom management. A deeper understanding of PD progression can help clinicians determine the most appropriate and personalized treatment for the patient. This can improve symptom control, allowing patients to maintain an optimal level of functioning for as long as possible.
      \item To develop new treatments. Understanding the progression of PD can help researchers identify new therapeutic targets and develop new treatments. This may lead to the discovery of new drugs and therapies to improve patient's quality of life and slow the progression of the disease.
     \item To provide information to patients and their families. Understanding the progression of PD can help patients and their families prepare for what may happen and take preventative steps to reduce the risk of complications. It can also help patients manage their expectations about the course of their disease and make informed decisions about their treatment and care.
\end{itemize}

The knowledge of the precise sequence of events in the progression of PD is important for:

\begin{itemize}
    \item Reducing heterogeneity in clinical trials;
    \item Monitoring the outcome of the treatment for new therapeutic interventions;
    \item Providing important information about the degeneration mechanism of PD.
\end{itemize}

This paper proposes a methodology to classify the different stages of PD. To date, several AI tools have been developed to diagnose the illness; none of them, however, have been created to determine the disease's progression stage. This paper aims to cover this gap by creating a model to determine the degree of progression using the "Parkinson's Progression Markers Initiative" (PPMI) dataset, which provides the patient's MRI and an indicator of the illness stage. 
We experiment with deep neural network models trained using the PPMI dataset containing MRI images, relating to several follow-up visits over time for the same patients. The dataset associates also each MRI image with the Hoehn and Yahr (H\&Y) scale~\cite{hoehn1967parkinsonism}, an internationally used PD progression rating method for clinical practice providing a score of the Parkinson's disease severity. We used them as a label for MRI images~\cite{templeton2022classification}.

Different pre-processing techniques have been applied to the MRI images in the dataset. Subsequently, to have sufficient elements to train a neural network, new images were generated through a data augmentation process.
The experimented neural network was composed of a 3D-CNN to extract the characteristics of the three-dimensional images, to which we added one or more recurrent layers (GRU or LSTM) to analyze the temporal data.

The paper is structured as follows:
Section~\ref{sec:related} illustrates the studies and works related to the proposed research. Section~\ref{sec:dataset} describes the Parkinson's Progression Markers Initiative dataset, and how it was used to derive the dataset used in this study.
Section~\ref{sec:methodology} presents the data processing and augmentation techniques, and describes the selected neural network models, while Section~\ref{results} discusses the results.
Finally, Section~\ref{sec:conclusion} concludes and presents future developments of this work.

\section*{Parkinson's Disease Background}
\label{sec:background PD}
Early recognition of PD is of great importance and can positively influence the patient's treatment and quality of life. Recognizing PD at an early stage allows treatment to be started promptly. While there is no definitive cure, there are pharmacological and non-pharmacological therapies that can help manage symptoms and improve quality of life. Some drugs used to treat PD are more effective in the early stages of the disease. Therefore, identifying the condition at an early stage can increase the likelihood of a positive response to treatment.
Furthermore, more effective symptom management strategies can be applied, for example through physical, occupational and speech therapies. These therapies can help maintain mobility, improve communication, and address any cognitive challenges.

PD is a brain disease that causes involuntary or uncontrolled movements, such as tremors, rigidity, bradykinesia, and difficulty with balance and coordination. Symptoms usually begin gradually and worsen over time, affecting the ability to walk, talk, and carry out normal daily activities. Motor symptoms are usually accompanied by mental and behavioural changes, sleep problems, depression, memory difficulties and fatigue~\cite{truong2011bradley}.
Although anyone can be at risk of developing PD, some research suggests that it affects men more frequently than women. Age is a significant risk factor, with most people developing the disease after age 60. However, approximately 5-10\% of cases have an onset before the age of 50, and in some of these cases, the disease may be hereditary~\cite{ferguson2016early}.

The causes of the onset of PD are not yet completely known, but being a multifactorial disease, causes may include:

\begin{itemize}
    \item \emph{\textbf{Damage to nerve cells:}} In PD, there is deterioration and sometimes death of nerve cells in the part of the brain called the basal ganglion, an area involved in the control of movements~\cite{braak2000pathoanatomy}.
    \item \emph{\textbf{Dopamine deficiency:}} Nerve cells in the basal ganglion normally produce a chemical called dopamine, but in PD the death or damage of these cells causes a deficiency of dopamine.
    \item \emph{\textbf{Presence of Lewy body:}} Many patients develop abnormal aggregates of proteins called Lewy bodies in their brain cells. These aggregates can interfere with the normal function of nerve cells~\cite{scatton1982dopamine}.
    \item \emph{\textbf{Genetic factors:}} While most cases of PD are not hereditary, there are cases in which the disease is linked to specific genetic variants. Numerous genes, including SNCA, Parkin, UCHL1, DJ-1, PINK-1, LRRK2, and ATP13A2, have mutations that cause familial forms of PD~\cite{rewar2015systematic}. However, genetics appears to play a limited role in most cases. 
    \item \emph{\textbf{Environmental factors:}} There are indications that exposure to certain environmental factors could increase the risk of developing Parkinson's disease. These factors may include exposure to environmental toxins or certain chemicals~\cite{priyadarshi2001environmental}.
\end{itemize}

There are currently no specific laboratory tests to diagnose PD, and the diagnosis is often based on the patient's medical history and neurological tests. If symptoms improve with medication, this may confirm the diagnosis. Some disorders can cause symptoms similar to PD, so it is important to rule out other possible causes~\cite{nutt2005diagnosis}.

While there is no cure for Parkinson's disease, there are treatments that can relieve symptoms. Drugs are often prescribed to increase dopamine levels in the brain or affect other neurotransmitters. Other therapies, such as deep brain stimulation, may be considered in severe or drug-resistant cases~\cite{cleary2015deep}.

PD is a complex condition that goes beyond its obvious motor symptoms. It has a profound impact on the psychological sphere of the individuals involved. The presence of debilitating motor symptoms, such as tremors and muscle stiffness, can trigger depressive and anxious moods. Awareness of disease progression and daily challenges can contribute to feelings of isolation and reduced psychological quality of life~\cite{dakof1986parkinson}.

Quality of life becomes a struggle for independence and social participation. The loss of autonomy in movement is more than a physical challenge; it's a psychological battle that can lead to feelings of frustration and isolation, and the psychological aspect plays a crucial role in how people cope with this disease. Cognitive-behavioral therapy can be an integral part of the treatment process. These approaches not only address specific psychological issues but also promote resilience and overall emotional well-being.

\subsection*{Hoehn and Yahr Scale}
\label{sec:scale}
 The most widely used tool for tracking PD progression, although combined with other clinical evaluations, is the Hoehn and Yahr scale (H\&Y Scale~\cite{hoehn1967parkinsonism},) ~\cite{bhidayasiri2012parkinson}. This scale offers a framework for evaluating PD development, which is essential for clinical care and to get a clear understanding of how the disease affects various facets of everyday life. 
The proposed stages of the Hoehn and Yahr Scale are:

\begin{itemize}
    \item \emph{\textbf{Initial stage (1-2).}} In the early stages, symptoms are often mild and may affect only one side of the body (unilateral). In \emph{\textbf{stage 1}}, symptoms are subtle and may not significantly affect daily life. In \emph{\textbf{stage 2}}, bilateral involvement begins to appear, but the posture is still preserved.
    \item \emph{\textbf{Intermediate stages (3-4).}} In \emph{\textbf{stage 3}}, impaired posture becomes evident, and the patient may begin to experience occasional falls. In the \emph{\textbf{stage 4}}, symptoms are more severe, with greater dependence on assistance to carry out daily activities. Posture is significantly compromised.
    \item \emph{\textbf{Advanced Stage (5).}} In the \emph{\textbf{stage 5}} and final stage, the disease reaches an advanced point with severe motor symptoms. Dependence on assistance is maximum, and the patient may require full-time care. Walking without assistance becomes impossible.
\end{itemize}

However, Hoehn and Yahr's scale only considers the motor symptoms of PD, ignoring several non-motor symptoms and psychosocial factors that can have a substantial impact on patients' quality of life.
Additionally, as Parkinson's symptoms may be heterogeneous, each person's rate of progression might be different. The scale provides a broad overview of the intensity of the motor symptoms, but effective clinical therapy necessitates a more thorough evaluation of non-motor symptoms, the overall course of the illness, and the influence on the patient's day-to-day activities.

\section*{Deep Learning Architectures: A Review}
\label{sec:background ML}

\subsection{Convolutional Neural Network}
\label{CNN}
Convolutional Neural Networks (CNNs) are a class of deep neural networks that are commonly used in image analysis and recognition \cite{sajjad2019multi}. They are particularly good at tasks like image classification, object detection and segmentation, image filtering, and feature extraction. Convolution layers take as input either an image or the output of other layers and they create feature maps. To do this, a kernel that is automatically trained by the network is used. In general, CNNs have layers organized hierarchically, so that the deeper the layer the more precise the attributes manipulated by its architecture. 

A 3D CNN extends the CNN concept to three-dimensional data, such as video data or medical imaging data with depth information that can be processed. Traditional CNNs operate on 2D data such as images, where convolutional operations are performed in two dimensions (height and width). In contrast, 3D CNNs can process spatial and temporal information simultaneously, making them suitable for tasks involving sequences of 3D data, such as video frames over time. This labour-intensive approach currently requires specialized analysis \cite{lu20193d,marwa2022mri}.

The application of 3D CNNs involves considering volumes of data rather than individual images. Every convolutional layer can capture and learn spatial and temporal patterns from the input data as well as execute three-dimensional convolution operations. This method has been demonstrated to be especially successful in situations where spatial depth or temporal sequence is critical, including tracing moving objects or making diagnoses based on medical imaging \cite{singh20203d}. The architecture of a 3D CNN includes:

\ \\
{\bf 3D Convolutional layers:}
\begin{equation*}
         y(i,j,k)= \sum_{l=0}^{L-1}\sum_{m=0}^{M-1}\sum_{n=0}^{N-1} x(i+l,j+m,k+n)*w(l,m,n) + b
\end{equation*}
where:
\begin{enumerate}
         \item $y(i,j,k)$ is the value of the resulting feature map at the location $(i,j,k)$.
         \item $x(i+l,j+m,k+n)*w(l,m,n)$ are the values of the input at the location $(i+l,j+m,k+n)*w(l,m,n)$.
         \item $w(l,m,n)$ are the weights of the convolution kernel.
         \item $b$ is the bias.
\end{enumerate}

\ \\
{\bf Pooling layers:}
\begin{equation*}
        y(i,j,k)= pooling function(x_{i,j,k}, x_{i+1,j,k},...,x_{i+P-1,j+Q,k+R-1})
\end{equation*}
where:
\begin{enumerate}
        \item $y(i,j,k)$ is the resulting value of the pooling layer,
        \item $P, Q, R$ is the size of the pooling window,
        \item $x_{i,j,k}$ are the values of the input in the pooling region.
\end{enumerate}

\ \\
{\bf Fully connected layers:}
\begin{equation*}
y= \sigma \left(\sum_{i=1}^{N}w_{i}*x_{i}+ b\right)
\end{equation*}
where:
\begin{enumerate}
        \item $y$ is the output of the fully connected layer,
        \item $x_{i}$ they are the outputs of the previous units,
        \item $w_{i}$ are the weights associated with each input,
        \item $b$: is the bias,
        \item $\sigma$ is the activation function.
\end{enumerate}
    
These formulas describe the fundamental operations in 3D-CNN, namely 3D convolution for feature extraction, 3D pooling for downsampling and fully connected layers for final classification. Convolutional operations are performed in three dimensions (width, height, and depth) to capture spatio-temporal features.

\subsection*{Recurrent Neural Network}
\label{RNN}
Recurrent Neural Networks (RNN) are suitable for sequential data or time series data ~\cite{zhang2019explainable}. These networks are commonly used for ordinary or temporal problems, such as translation, language processing (NLP), speech recognition, and image subtitling.
Like CNN, RNNs use training data to learn. Their importance relies on the "memory" property as the output of the current input depends on past computations ~\cite{kriegeskorte2019neural}. 
While traditional deep neural networks assume that inputs and outputs are independent, the output of RNNs depends on the preceding elements within the sequence \footnote{https://www.ibm.com/it-it/topics/recurrent-neural-networks}. 
RNNs can also be stacked and used to learn temporal information in both directions, and in this case, we refer to them as bidirectional RNNs (BRNNs). Their most important property consists in being able to make predictions using both past and future data ~\cite{he2021db}.

Among RNNs, there are two very popular architectures: the Long Short-Term Memory (LSTM) and the Gated Recurrent Unit (GRU). 

\subsubsection*{Long Short-Term Memory}

LSTMs were created in response to the issue of gradient disappearance in conventional neural networks. Gradient disappearance can happen when a neural network's inability to learn long-term associations is caused by incorrect backpropagation during training, which results in very small gradients. LSTMs were introduced to address this issue and enable neural networks to record long-range temporal dependencies ~\cite{shewalkar2019performance}.

\begin{figure}[ht]
	\centering
	\includegraphics[width=0.8\textwidth]{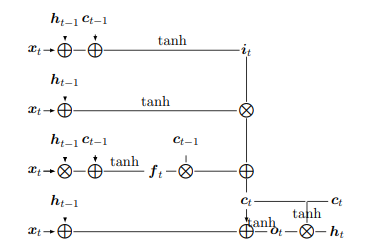}
	\caption{An example of an LSTM architecture}
	\label{fig:LSTM}
\end{figure}

They introduce the concept of "gate", which allows for control of the flow of information necessary to predict the output in the network through operations on matrices. There are three types of gates in LSTMs as shown in Fig.~\ref{fig:LSTM}.

\ \\
{\bf Input Gate:} it decides which new information will be stored in the memory cell, and it is described by:
\begin{equation*}
         i_{t} = \sigma (W_{ii} * x_{t} + b_{ii} + W_{hi} * h_{t-1} + b_{hi})
\end{equation*}
\begin{equation*}
        g_{t} = \tanh (W_{ig} * x_{t} + b_{ig} + W_{hg} * h_{t-1} + b_{hg})
\end{equation*}
\begin{equation*}
        a_{t} = i_{t} * g_{t}
\end{equation*}

\ \\
{\bf Forget Gate:} it decides which information in the memory cell should be deleted or forgotten, and it is characterized by the following relations:

\begin{equation*}
         f_{t} = \sigma (W_{if} * x_{t} + b_{if} + W_{hf} * h_{t-1} + b_{hf})
\end{equation*}
\begin{equation*}
        o_{t} = \tanh (W_{og} * x_{t} + b_{og} + W_{hg} * h_{t-1} + b_{hg})
\end{equation*}
\begin{equation*}
        a_{t} = f_{t} * g_{t-1}
\end{equation*}

\ \\
{\bf Output Gate:} it calculates the output based on the current memory cell and the current input through the following expressions:
\begin{equation*}
         o_{t} = \sigma (W_{io} * x_{t} + b_{io} + W_{ho} * h_{t-1} + b_{ho})
\end{equation*}
\begin{equation*}
        h_{t} = o_{t} * \tanh (c_{t})
\end{equation*}

Furthermore, in LSTMs, we also have \emph{\textbf{Cell State Update}} which is responsible for updating the cell state based on the information received from the input and forget gates. The formula for updating the cell state in LSTMs is as follows:

\begin{equation*}
c_{t} = f_{t} * c_{t-1} + a_{t}
\end{equation*}
where the different symbols indicate:
\begin{itemize}
    \item $x_{t}$ is the current input.,
    \item $h_{t-1}$ is the output of the LSTM cell at the previous time step, 
    \item $W$ and $b$ are the weights and biases of the model,
    \item $\sigma$ is the sigmoid function,
    \item $\tanh$ is the hyperbolic tangent function.
\end{itemize}

These formulas describe the flow of information through an LSTM cell. The cell stores a state $c_{t}$ which is updated or forgotten based on the input, forget, and output ports. The output $h_{t}$ is then calculated based on the information in the cell state, and then filtered through the output gate \cite{gao2016deep}.
     
\subsubsection*{Gated Recurrent Units} 
\begin{figure}[ht]
	\centering
	\includegraphics[width=0.7\textwidth]{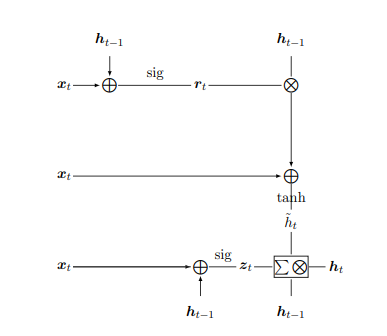}
	\caption{An example of a GRU architecture}
	\label{fig:GRU}
\end{figure}

GRUs are another variant of RNN designed to address the limitations of traditional neural networks in modelling complex temporal sequences. GRUs are structured in a simpler way than LSTMs \cite{can2020gating} as they use only two gates as shown in Fig.~\ref{fig:GRU}:

\ \\
{\bf Reset Gate:} This gate decides how much of past information should be erased and it controls how much of past information should be mixed with new information.
\begin{equation*}
        z_{t} = \sigma (W_{z} * x_{t} + U_{t} * h_{t-1} + b_{z})
\end{equation*}

\ \\
{\bf Update Gate:} This gate determines how much past information should be added to the existing information stored in the memory cell.

\begin{equation*}
        r_{t} = \sigma (W_{r} * x_{t} + U_{r} * h_{t-1} + b_{r})
\end{equation*}

In GRUs, we also have the notion of \emph{\textbf{Candidate Hidden State}} and it is calculated as a weighted combination between the output of a reset gate $(r_{t})$ and the previous hidden state $h_{t-1}$, along with the current input $x_{t}$, using the following formula:

\begin{equation*}
    \overline{h_{t}} = tanh (W * x_{t} + U * (r_{t}\odot h_{t-1}) + b)
\end{equation*}

Then, the \emph{\textbf{Hidden State Update}} is performed using a weighted update operation between the candidate hidden state and the previous hidden state. Below is the formula:

\begin{equation*}
    h_{t} = (1 - z_{t})\odot h_{t-1} + z_{t}\odot \overline{h_{t}}
\end{equation*}

The different symbols indicate:

\begin{itemize}
    \item $c_{t}$ is the state of the cell at the time $t$,
    \item $f_{t}$ is the output of the forget gate, 
    \item $c_{t-1}$ is the state of the cell at the previous time $t-1$,
    \item $i_{t}$ is the output of the input gate,
    \item $g_{t}$ is the output of the candidate gate,
    \item $W,U,b$ are the weights and biases of the model,
    \item $\odot$ denotes the element-wise product
\end{itemize}

These formulas describe how the candidate hidden state $\overline{h_{t}}$ is calculated and how the hidden state $ h_{t}$ is updated using information from the hidden state candidate and the update gate $z_{t}$. The update gate governs how much of the new candidate's hidden state should replace the previous hidden state.

Unlike LSTMs, GRUs do not have a separate forget gate. The lack of this gate simplifies the structure of GRUs, making them faster to be trained and less computationally expensive than LSTMs.

\section*{Related Works}
\label{sec:related}
Extensive research has been conducted for a long time in the medical field on clinical, cognitive, and biomarker neuroimaging datasets for normal ageing and Parkinson's disease \cite{saravanan2022systematic}.  
Several models and techniques of Artificial Intelligence have been applied to the medical field and in particular, in the area of neuroimaging for detecting PD. %However, few of them are related to the Parkinson's case. 
In this section, we discuss related work adopting AI techniques for supporting both PD diagnosis and classification of the progression level.

Vyas et al.~\cite{vyas2022deep} used MRI images to diagnose PD. They also identified biomarkers that can be used to observe how PD spreads. A total of 318 MRI scans from the Parkinson's Progressions Markers Initiative (PPMI) were utilized to create the dataset and to train and test two models: a 2D CNN and a 3D CNN. the 3D model is more accurate, reaching 88.9\% classification accuracy. 

Butt et al.~\cite{butt2018objective} assessed the capacity of the Leap Motion Controller, a device able to measure positional data, to detect PD by evaluating the motor impairment of patients.  To this aim, they recruited 16 subjects with Parkinson's disease and 12 healthy individuals and performed data acquisition from the leap motion of both samples. They assessed the performance of three classification models: logistic regression, Naïve Bayes, and Support Vector Machine. The best performance was reached by logistic regression (70.37\% accuracy).

Templeton et al.~\cite{templeton2022classification} adopted Machine Learning (ML) to detect the PD level on the basis of digital neurocognitive assessments (e.g., digital biomarkers) consisting of several tests on a tablet. 
In this study, 75 participants — 50 people with PD and 25 controls — took part in functional movement evaluations, standardized health questionnaires, and 14 tablet-based neurocognitive functioning tests. The PD patients are classified according to the H\&Y scale. Decision tree classification of sensor-based characteristics enabled the differentiation of motor dysfunction in patients with PD from healthy controls with an accuracy of 92.6\% and the early and advanced phases of PD with an accuracy of 73.70\%.

Ferreira et al.~\cite{ferreira2022machine} performed gait analysis to support both the PD detection and the classification of the PD phases considering spatio-temporal parameters of walking features, such as variability and asymmetry using wearable sensors. 42 people with mild and 21 people with moderate PD make up the dataset. The tests examined gait data collected from PD patients with varying degrees of PD motor symptom severity and 63 matched control group participants. The authors performed the classification between PD and non-PD patients reaching an accuracy rate of 84.6\% with the Naïve Bayes algorithm. The progression level has been measured by adopting the H\&Y classification. The best PD progression classification performance is reached by the Random Forest algorithm, by reaching an average Area Under the ROC curve of 0.786. In our case, with an RMI analysis, we reached an average Area Under the ROC curve of 91.90\%.
 
Veetil et al.~\cite{veetil2021parkinson} applied and evaluated the use of pre-trained models AlexNet, Xception, DenseNet201, ResNet50, VGG16 and VGG19 for Parkinson's classification. They also adopted the PPMI dataset. Using pre-trained models with images in two dimensions from the ImageNet dataset, VGG19 obtained an accuracy of 92.6\%.
%Techniques for generating medical images were used to improve model performance in another study. In this way, the problem of the limited quantity of data present in the dataset was solved and allowed to improve the generalization capacity of the trained model\cite{duran2021ordinal}.

In~\cite{prashanth2016high} 123I-ioflupane thermal brain (SPECT) and 123I-ioflupane positron emission tomography (PET) images were taken from PPMI and analyzed to discriminate Parkinson's patients from controls. The features selected to train linear (Linear Regression, Quadratic Discriminant Analysis) and nonlinear (Support Vector Machine - SVM, Random Forest, neural network) classifiers were shape indices of the segmented contours (ellipticity, compactness, roughness, etc.). SVM  achieved 97\% accuracy. The participants in the study were 208 healthy normals, 427 PD, and 80 SWEDD, i.e., individuals with mutation of the DYT11 gene in patients without dopaminergic deficiency.

Adeli et al.~\cite{adeli2016joint} proposed the Joint Feature-Sample Selection (JFSS) method, a classification framework for simultaneously decluttering the selected subset of features and samples and learning a classification model. To evaluate the proposed method, they used the PPMI and ADNI datasets. They considered T1 MRI collections of Parkinson's patients and controls, by extracting volumetric and shape characteristics from different brain regions. Data were normalized to reduce anatomic-functional variations. Noise and outliers were removed. The selected features were used to train linear and nonlinear classifiers. Results indicated that the proposed method can identify imaging biomarkers and diagnose the disease with an accuracy of 81.9\% using robust Linear Discriminant Analysis.

Peng et al.\cite{peng2017multilevel} proposed a multilevel Region-Of-Interest feature-based machine learning method that uses brain T1 MRI images from the PPMI dataset (69 PD patients and 103 normal controls) to detect PD-related morphometric biomarkers and improve disease identification power. The feature selection method based on filters and wrappers and the multi-kernel SVM are used in the classification algorithm and achieved an accuracy of 85.78\%.

In~\cite{shahid2020deep} the voice of PD patients is analyzed to predict the Motor and Total-UPDRS (Unified Parkinson's Disease Rating Scale) scores. The voice measurements were captured automatically from the patient’s home for 6 months. The dataset used was taken from the machine learning repository provided by the University of California at Irvine (UCI). Multicollinearity problems in the data set are addressed and the input feature space is made smaller using principal component analysis (PCA). They applied a neural network to the condensed input feature space, achieving for the Motor-UPDRS a coefficient of determination 97\% and 94.8\% in the test and validation phases, respectively. Concerning the Total-UPDRS,  the coefficient of determination was 95.6\% and 92.9\% in the test and validation phases, respectively.

In \cite{dadu2022identification} Dadu et al. used machine learning methods (Random forest, LightGBM, XGBoost, SVM, Clustering and Lasso-regression models) on complete clinic longitudinal data from the PPMI (n = 294 cases) to identify patient subtypes and predict the progression of the illness. The resulting models were validated in an independent, well-defined clinical panel (n = 263 cases). The analysis identified three distinct subtypes, which correspond to a slow, moderate, and rapid progression of the illness. Five years after the initial diagnosis, they obtained extremely accurate predictions of the malady's progression, with AUC values of 0,92 for the group experiencing the slower progression, 0,87 for the moderate advancement, and 0,95 for the group experiencing the faster progression.

García-Ordás et al. \cite{garcia2023determining} proposed a method based on deep learning techniques for early diagnosis and severity prediction of Parkinson's disease through voice analysis. The dataset used consists of data extracted from the speech analysis of patients, with the use of an autoencoder to intelligently reduce the features. The work is divided into two steps: i) classification to determine the severity of Parkinson's disease; and ii) regression to predict the degree of disease evolution through UPDRS. The proposed model is a mixed multi-layer perceptron, capable of performing classification and regression simultaneously. The results show a success rate of 99.15\% in classifying severity and a root mean square error of 0.15 in predicting the degree of disease involvement.

Severson et al.\cite{severson2021discovery} proposed a study that focuses on creating a statistical progression model for Parkinson's disease that takes into account intra-individual and inter-individual variability and drug effects. Using longitudinal data from 423 early-stage Parkinson's patients and 196 healthy controls collected for up to 7 years, the model identifies eight disease states based on motor and non-motor symptoms. The most notable finding is the discovery of overlapping and non-sequential progression trajectories, challenging the assumption of deterministic progression patterns across disease subclasses. This approach provides a comprehensive view of PD progression, highlighting the heterogeneous nature of the disease and suggesting that static subtype assignment may not be effective in capturing the full spectrum of PD progression. The model could have significant implications in the management of PD patients and the design of clinical trials, serving as a prognostic and predictive tool.

Most of the works proposed in the literature deal with recognising Parkinson's. However, it is also important to identify the level of PD. For this reason, we propose an approach to identify the level of progression of Parkinson's disease in patients by examining their MRIs. Determining disease staging is crucial because it can be used to predict future illness development and identify people who might benefit from early therapeutic interventions.

This work differs from the others because we consider a time series of MRI images to train a multi-class classifier. 
Thus, this classifier learns to identify the differences between images of different PD levels according to the H\&Y scale. We use RNN+CNN hybrid neural networks. The first network during the training examines the images of the time series while the second analyzes each single image in detail.

\section*{The Dataset}
\label{sec:dataset}
The dataset adopted is the "Parkinson's Progression Markers Initiative" (PPMI)~\cite{marek2011parkinson}, launched in 2010 to identify biomarkers of the onset and progression of Parkinson's disease. 
PPMI is mainly sponsored by "The Michael J. Fox Foundation for Parkinson's Research", a foundation created in 2000 by the homonymous actor Michael J. Fox, who suffers from this disease, in collaboration with researchers, funders, governments, medical centres in the United States, Europe, Israel and Australia.
33 clinical centers in 11 countries take part in this initiative which collects data and biological samples of 1,400 participants, constituting the most robust Parkinson’s database and sample bank ever created.
This comprehensive and standardized repository of Parkinson's disease data is a biological sample repository and is coupled with a simple online system that allows researchers to access study data and samples for complementary disease research in their laboratories across the country. world. 
Through the platform provided by the PPMI, it is possible to access a lot of information and data, some of a structural-descriptive nature that is used to understand the structure of the dataset and the semantics of the attributes of the individual tables and others containing information regarding specific medical examinations: from the detected values from a simple blood sample to the responses of neuro-cognitive questionnaires up to the results of diagnostic tests, all collected in the form of tables and placed in specific reference areas that can be classified according to the type of study conducted on the patient.
The tables contain patient data that can be classified into seven macro-areas according to the type of information collected:

\begin{enumerate}
    \item \emph{Biospecimen:} collection of data relating to clinical tests such as blood sampling, DNA, lumbar puncture;
    \item \emph{Imaging:} I tell about relevant data through the use of imaging techniques, such as Magnetic Resonance and DatScan;
    \item \emph{Medical History:} clinical history of patients from the first symptoms of the disease to the latest health conditions. The collection includes any side effects of medications taken, neurological, and physical test results, and so on.
    \item \emph{Motor MDS – UPDRS:} collection of data relating to motor disorders using the MDS-UPDRS scale to assess the stage of Parkinson's disease.
    \item \emph{Non-Motor Assessments:} area that collects data relating to cognitive and emotional behavioural disorders.
    \item \emph{Study Enrollment:} area includes the conclusive data of particular studies conducted on patients.
    \item \emph{Found:} area that includes various questionnaires regarding patients' habits and lifestyles.
\end{enumerate}

In this study, the data present in the "Imaging" macro-area will be used to acquire the medical images necessary for this research and the data present within the "Motor MDS-UPDRS" macro-area to assign a score to each image of the dataset.

The MDS-UPDRS scale consists of four parts:

\begin{enumerate}
   \item \emph{Part 1:} Non-motor aspects of daily life. It is divided into two parts:
    \begin{itemize}
        \item \emph{Section 1A} deals with some behavioural aspects, assessed by the examiner based on the relevant %information obtained from the patient and the caregiver;
        \item \emph{Section 1B} is completed by the patient with or without the help of the caregiver and independently of the researcher. The questions posed to the patient concern: sleep disturbances, urinary or constipation problems, apathy, hallucinations and psychosis.
    \end{itemize}
    \item \emph{Part 2:} Motor aspects of daily life. Like Part 1 Section 1B, this questionnaire is completed by the patient with or without the help of the caregiver but can be reviewed by the researcher to verify its completeness and clarity. The questions include: writing, speaking, salivation, tremor, chewing and motor block;
    \item \emph{Part 3:} Motor Exam. This part deals with the evaluation of the motor signs of Parkinson's present in the patient. The examiner shows the patient the task that (s)he must complete and in this phase, the examiner "evaluates what he sees". This analysis includes leg agility, postural stability, rigidity, quality of movement in the act of getting up from the chair, repeated movements of the fingers and kinetic tremor of the hand;
    \item \emph{Part 4:} Motor Complications. This part includes both instructions for the examiner and instructions to read to the patient. It is filled in by the examiner who is responsible for integrating the information obtained from the patient with his evaluation and observation. The examiner uses both anamnestic and objective information to evaluate two motor complications, dyskinesias (random involuntary movements) and motor fluctuations.
\end{enumerate}
  
The questionnaires are filled in each visit and a score is assigned which is a numerical index indicating the progress of the disease. The higher this value is, the more advanced the disease is.
The visits are classified and ordered in the following way:

\begin{itemize}
    \item \emph{SC:} Screening Visit, a preliminary visit lasting approximately eight hours, aiming at assessing whether the patient may participate in the project;
    \item \emph{BL:} Baseline Visit, the visit in which the patient's condition was first assessed (before he/she had received any treatment);
    \item \emph{V01-V15:} sequence of tests that have been programmed for all patients;
    \item \emph{ST:} Symptomatic Therapy, treatment aimed at alleviating the symptoms of the disease;
    \item \emph{LOG:} visit aimed at recording adverse events.
\end{itemize}

The data adopted in this study were extracted from the table related to two macro-areas: Imaging, containing the medical images (Magnetic Resonance and DatScan), and an index describing the progression of PD, calculated according to the values of the H\&Y scale. This scale describes five stages in the progression of Parkinson's disease. Like in~\cite{butt2018objective} we also adopt the H\&Y Scale (mentioned in \ref{sec:scale}) for labelling the RMI images for supervised learning tasks.

\subsection{Dataset creation}
Before introducing the dataset it is important to point out that there exist two types of MRI: T1-weighted and T2-weighted images. 
In MRIs, the magnetically sensitive hydrogen nuclei align with the fields created by the MRI. The radio waves released by the scanner are then absorbed as energy by the nuclei.
When the radio waves stop, the nuclei return to their initial state, emitting tiny energy signals that the computer uses to create visual images. This signal has two components:

\begin{enumerate}
  \item Longitudinal
  \item Transversal
\end{enumerate}
 
Initially, the longitudinal signal is weak because most nuclei are tilted to the axis. However, this signal grows as the nuclei realign. The time constant that determines the realignment speed is denoted T1.
Conversely, the transverse signal is strong because most nuclei are in phase coherence. The signal decays as the nuclei go out of phase as they realign. This decay is indicated by the time constant T2\cite{bolas2010basic}.
The differences in T1 and T2 relaxation times control the contrast of the MRI and depending on the case their values are modelled to have more or less contrasted images.

We focus on a single type of MRI and we selected T2-weighted images because the contrast and brightness are predominantly determined by the T2 properties of tissue~\cite{greve2009accurate}.

Images of patients were selected with the following characteristics:

\begin{enumerate}
    \item The patient must have Parkinson's disease;
    \item There must be at least two MRI scans acquired in two different visits;
    \item MRIs must be T2w.
\end{enumerate}
 
The first two constraints are dictated by the formulation of the problem, as we want to recognize the progress of PD in a series of images. 

Following these specifications, 232 MRI scans were collected related to the brains of 63 patients with Parkinson's disease.

For classifying the level of the patient progression~\cite{templeton2022classification}, in our studio, five classes were identified according to the H\&Y scale. Then, noting that the fourth and fifth classes in the PPMI dataset had few samples, the patients of these two classes were merged into a single class. 

The classes correspond to the following symptoms:

\begin{itemize}
    \item \emph{Class 1:} No signs of disease
    \item \emph{Class 2:} Unilateral disease
    \item \emph{Class 3:} Bilateral disease, without impairment of balance and mild to moderate bilateral disease, some postural instability, physically independent
    \item \emph{Class 4:} Severe disability, still able to walk or stand unassisted and wheelchair-bound or bedridden unless aided or needing a wheelchair or bedridden unless assisted.
\end{itemize}

The selected images are grouped into classes according to this scale. The cardinality of the classes is reported in Table~\ref{tab:classes}.

\begin{table}[h!]
	\caption{MRI classes}
	\centering
	\resizebox{0.6\columnwidth}{!}{
		\begin{tabular}{|c|c|}
		\hline
			\textbf{ID Class} & \textbf{Number of elements} \\
			\hline
			1 & 43 \\
			\hline
			2 & 124 \\
			\hline
			3 & 42 \\
			\hline
			4 & 22 \\
			\hline
		\end{tabular}}
	\label{tab:classes}
\end{table}

\section*{Methodology}
\label{sec:methodology}
\begin{figure*}[ht]
	\centering
	\includegraphics[width=1\textwidth]{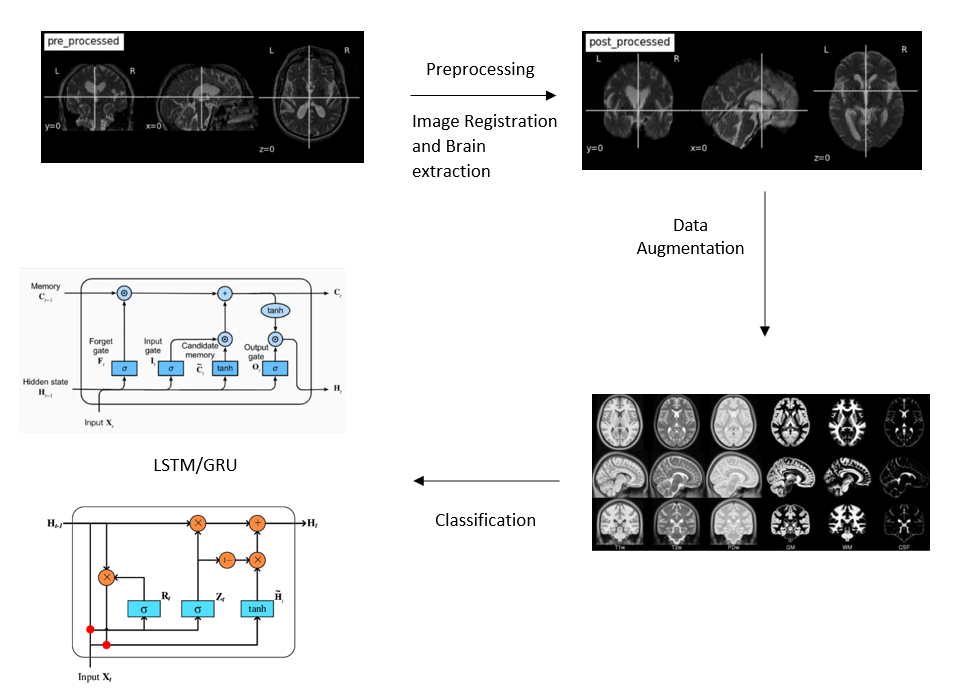}
	\caption{Overview of the proposed method.}
	\label{fig:process}
\end{figure*}

Here we describe the proposed method for designing a classifier of the level of severity of PD using MRI images extracted from the PPMI dataset. An overview of the proposed method is depicted in Fig.~\ref{fig:process}.
It is articulated in three phases: data preprocessing,
data augmentation, and classification, as detailed in the following.

\subsection*{Data Preprocessing}
\label{sec:preprocessing}
The correction and cleaning of the MRIs is an essential process, as this type of image could present different problems, mainly due to the acquisition method. All the pre-processing techniques were performed by using Nipype~\cite{gorgolewski2011nipype}. This Python project offers a standard interface to current neuroimaging programs and makes it easier for these programs to interact with one another inside a single workflow.
The MRIs in the PPMI dataset are three-dimensional: this means that their acquisition process is slightly different from the usual one. Generally, an MRI consists of an image in two dimensions. From it, it is possible to build an image in three dimensions. The acquisition takes a set of images that cover the entire surface useful to reconstruct a three-dimensional image.
This type of image complicates the analysis, as the patient could move during the acquisition, creating a misalignment between the various "slices" acquired~\cite{jones2010twenty}.

\subsubsection*{Conversion from DICOM to NIfTI}
\begin{figure*}[ht]
	\centering
	\includegraphics[width=0.6\textwidth]{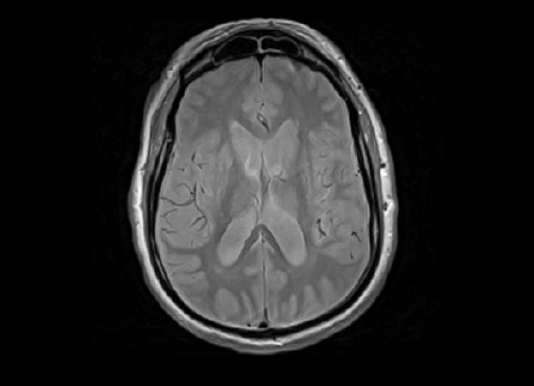}
	\caption{MircroDicom interface}
	\label{fig:fig_a}
\end{figure*}

\begin{figure*}[ht]
	\centering
	\includegraphics[width=0.6\textwidth]{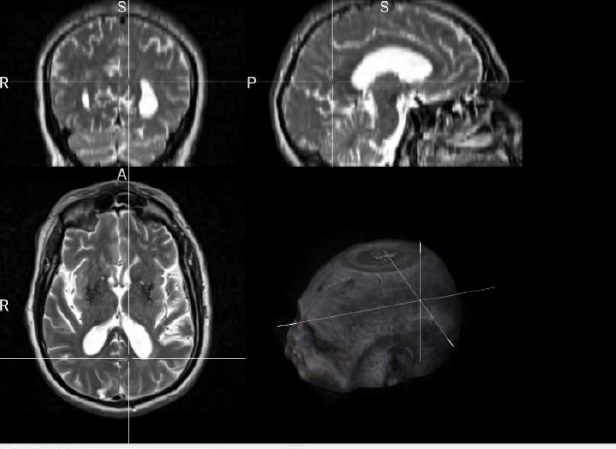}
	\caption{MRIcroGL interface}
	\label{fig:fig_b}
\end{figure*}

The PPMI dataset contains MRI images of patients in DICOM (Digital Imaging and Communications in Medicine) format files~\cite{modi2021deep}, which represents a two-dimensional image of a single angle of the person; therefore, the first step of the pre-processing was the reconstruction in a three-dimensional format, or in NIfTI (Neuroimaging Informatics Technology Initiative)~\cite{strother2006evaluating}, a format notoriously used in the medical field.
Fig \ref{fig:fig_a}(a) shows the original DICOM format 
 of an MRI belonging to PPMI visualized by using the tool “MicroDicom”~\cite{inthavong2021list}, while Fig.~\ref{fig:fig_b}(b) shows the three-dimensional NIfTI format if the same image produced by the “MRIcroGL”\cite{sharma2020automated} tool.

\subsubsection*{Image Registration} \label{imageregistration}
\begin{figure*}[ht]
	\centering
	\includegraphics[width=1\textwidth]{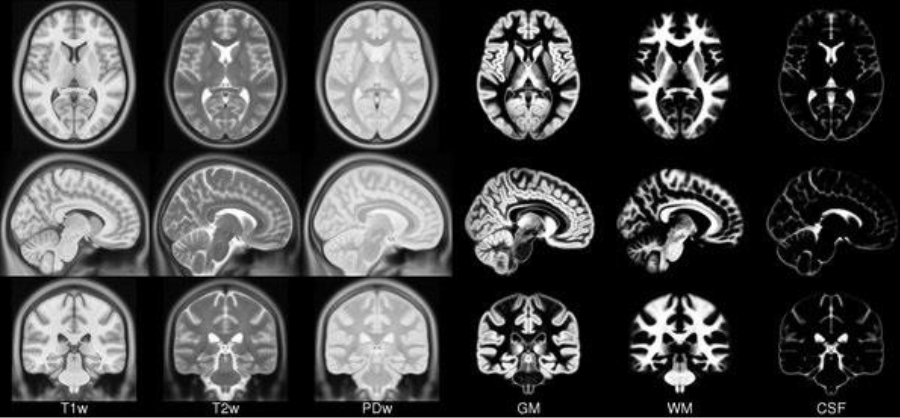}
	\caption{Example of templates.}
	\label{fig:template}
\end{figure*}

The image registration process consists of combining two images by superimposing one image to the other~\cite{mani2013survey}.
In the medical field, image registration allows you to combine data from multiple modalities to obtain comprehensive patient information: it can help monitor tumour growth, facilitate treatment verification, improve interventions, or compare patient data with a template.
To apply image registration we need a starting image, a registration algorithm, and a target image, in our case a template. A template is an MRI in a particular spatial location (see Fig. \ref{fig:template} as an example).
Model building is an iterative process that involves normalizing, aligning, and averaging an MRI set from different patients~\cite{forbes2021processing}.
Templates serve as a common reference space and allow researchers to combine and compare data from multiple people and play an important role in a variety of neuroimaging activities:

\begin{itemize}
    \item For the spatial normalization of the input image;
    \item For tissue segmentation;
    \item For the analysis of the regions of interest.
\end{itemize}

\begin{figure}[ht]
	\centering
	\includegraphics[width=0.7\textwidth]{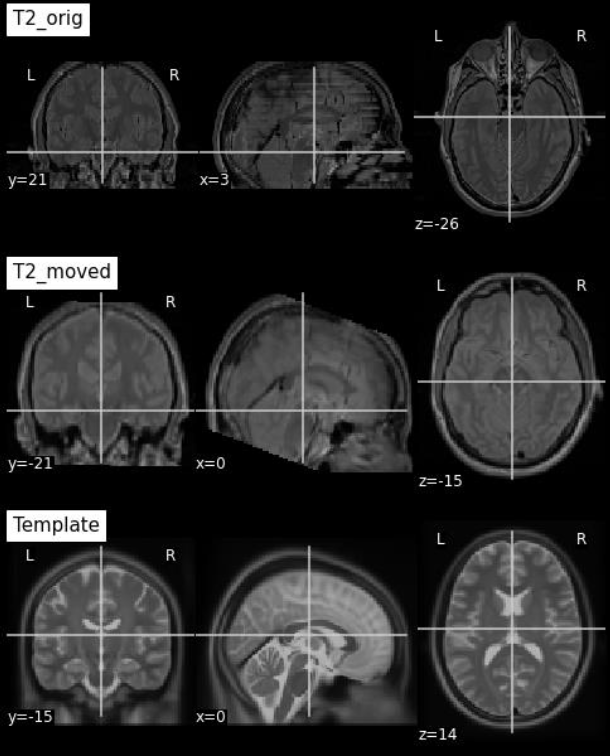}
	\caption{Image registration.}
	\label{fig:image_registration}
\end{figure}

Through the FLIRT~\footnote{\url{https://fsl.fmrib.ox.ac.uk/fsl/fslwiki/FLIRT}} tool, image registration was applied using the templates available on TemplateFlow~\cite{li2021moving}, as shown in Fig.~\ref{fig:image_registration}. 

\subsubsection*{Brain Extraction}
In an MRI image, in addition to sections of the brain, there are also non-brain tissues, such as fat, skull or neck, which can cause problems with the analysis~\cite{kharb2021review}. 
Brain extraction, also known as "skull-stripping", describes the process of removing the skull and non-brain tissues from brain MRI scans. it is a method for segmenting the brain from non-brain into structural and functional data.~\cite{pei2022general}.
It is also a necessary processing step in most studies for compliance with privacy regulations. 
For these reasons, several brain extraction algorithms (BEAs) are available in neuroradiological research \cite{igwe2022automatic}.
 
First, a mask is generated to identify brain tissues that include the grey matter, white matter, and cerebrospinal fluid of the cerebral cortex and subcortical structures, such as the brain stem and cerebellum. The scalp, hard matter, fat, skin, muscles, eyes, and bones are classified as non-brain tissues~\cite{ramasamy2021segmentation}.

\begin{figure*}[ht]
	\centering
	\includegraphics[width=1\textwidth]{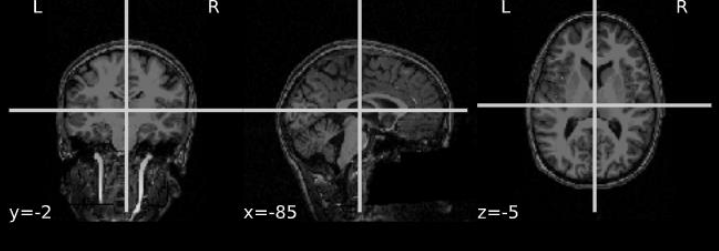}
	\caption{MRI before Brain Extraction.}
	\label{fig:brain_extraction_a}
\end{figure*}

\begin{figure*}[ht]
	\centering
	\includegraphics[width=1\textwidth]{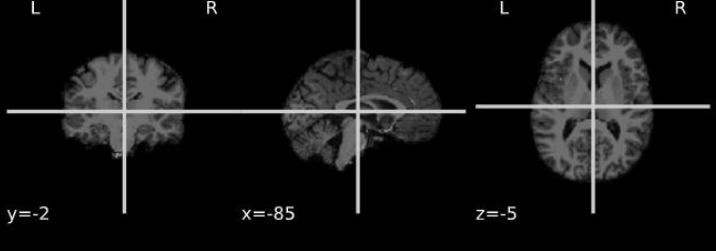}
	\caption{MRI after Brain Extraction}
	\label{fig:brain_extraction_b}
\end{figure*}

Brain extraction was applied by using the BET tool~\cite{smith2000bet}. An example of the result of brain extraction is shown in Figs.~\ref{fig:brain_extraction_a} \ref{fig:brain_extraction_b}.

\subsection*{Data Augmentation}
Data augmentation is a technique for creating new data from current ones to artificially increase the amount of data~\cite{chlap2021review}. 
It is widely used in many machine learning tasks, such as object detection, classification, and understanding of natural language. The creation of new and varied instances of training data may improve the functionality and results of deep learning models.
Data augmentation techniques may include making small changes to the data or using machine learning models to produce new data from the original ones.

Two types of data can be distinguished, namely:

\begin{itemize}
    \item \emph{Synthetic:} The data are created artificially, usually through generic networks without using elements of the real environment, but often data that do not comply are generated~\cite{bui2022virtual}.
    \item \emph{Generated:} These are data produced by geometric alterations (such as overturning, translation, rotation, or addition of noise) starting from original data~\cite{aderghal2017classification}~\cite{hao2021comprehensive}.
\end{itemize}
  
The dataset at our disposal consisting of 232 MRIs was completely unbalanced in the number of elements per class (see Table \ref{tab:classes}). However, the lack of MRI is a common problem in many studies that use artificial intelligence. 

At first, simple transformations such as rotation or flip were applied to images to obtain four balanced classes, but, even applying this type of transformation we did not obtain sufficient data for the training. Furthermore, this technique may have some problems: applying the same distortions to the entire MRI sequence of the same patient would maintain the relationship of the input data but in some cases, we could create unrealistic elements as in the case of rotations. F

One solution could have been to use generative networks (GAN)~\cite{rejusha2021artificial}.  However, we were not able to verify that the images maintained the same progression characteristics of Parkinson's disease among the various magnetic resonances of an individual. 

A study \cite{nalepa2019data} has addressed this problem using image registration, showing that the procedure offers a statistically significant increase in the performance of the model trained with the generated data.
To obtain realistic MRIs we adopted this methodology, already used in the image registration phase (subsection \ref{imageregistration}), which applies transformations to the input images by superimposing them to templates.
We applied this technique using different templates to create new elements starting from the current MRI.

We performed the registration related to data augmentation by using the Jupiter notebook offered by Google Colab. In particular,  we adopted the paralyzing feature of the Nipype interface (in Google Colab a dual-core machine is assigned), with the average time for processing each MRI pair in about 400 seconds.

The proposed techniques enabled us to create new elements, useful to properly train the neural network.
In particular, we adopted six templates. The number of images generated for each class depended on the need for balancing that class concerning the others.
For example, for class four, all six templates were used to generate enough images to balance the number of elements in class two. Instead, for class one two templates were used to match the number of elements in class two.
After creating the new images and balancing the classes, other elements were generated starting from the current ones using the flip transformation.
As a result of these processes, the final dataset distribution in classes is shown in Table~\ref{tab:classes2}. 

\begin{table}[h!]
	\caption{MRI classes after Data Augmentation}
	\centering
	\resizebox{0.6\columnwidth}{!}{
		\begin{tabular}{|c|c|}
		\hline
			\textbf{ID Class} & \textbf{Number of elements} \\
			\hline
			1 & 375 \\
			\hline
			2 & 375 \\
			\hline
			3 & 375 \\
			\hline
			4 & 375 \\
			\hline
		\end{tabular}}
	\label{tab:classes2}
\end{table}

\subsubsection*{Augmentated dataset's validation}
To validate the generated MRI images, a classifier trained on the MRI images of the dataset was used.
The data necessary for validation were always collected through the resources provided by the PPMI.
In total, 300 MRIs were collected, 150 MRIs from patients with Parkinson's disease and 150 MRIs from volunteers without the disease.
Before being used for training the network, the images were resized and another 600 MRIs were generated from them by applying the inversion transformation (both left and right)\cite{safdar2020comparative,badvza2020classification}. Finally, we used a neural network inspired by a study in which several convolutional networks were evaluated for MRI analysis of tuberculosis patients\cite{bakas2018identifying}, in particular, a convolutional network in three dimensions (CNN-3D)\ref{CNN}.
The dataset was divided into a training set and a validation set with a ratio of 70\% and 30\%. 
The training was carried out using the GPU provided by Google Colab. The neural network during the training achieved good results with an accuracy on the training set of 0.97\% and the validation set of 0.91\%. Finally, to test our network on the test set, about a hundred samples were taken from the augmented dataset, obtaining approximately 95\% accuracy.

\subsection*{Classification}
The goal of this work is to classify Parkinson's progression level using MRI images.  To this aim, we trained the Deep Learning classifier with ordered sequences of MRIs of the same patient, labelled by the class of each image. We decided to use mixed neural networks, in particular, a 3D-CNN\ref{CNN} having as the last layers of an RNN\ref{RNN}.

In fact, in our case, we cannot send an MRI sequence directly to an RNN for classification but, first, we need to extract the characteristics of the images and then use the latter for temporal analysis of the data~\cite{dubey2011evaluation}. For the extraction of characteristics, it is possible to use a CNN, which receives an image and applies the convolution to find its characteristics to classify it.
By joining these two networks, we will have a network that will be able to extract the characteristics of the images and analyze their temporal behaviour. 

Because we provide input for several MRIs to the first level, we need to avoid merging them. Therefore, for each image taken in input, we will apply the convolution and then we will continue the classification process with other layers that will be in common with all the images.

\begin{table}[h!]
	\caption{CNN-3D Layers of 3DCNN+LSTM/GRU architecture.}
	\centering
	\resizebox{0.7\columnwidth}{!}{
		\begin{tabular}{|c|c|c|}
		\hline
			\textbf{Layer (type)} & \textbf{Output Shape}  & \textbf{Param} \\
			\hline
			InputLayer & [(None, 128, 128, 64, 1)] & 0 \\
			\hline
			Conv3D & (None, 126, 126, 62, 64) & 1792\\
			\hline
 		MaxPooling3D & (None, 63, 63, 31, 64) & 0\\
			\hline
			BatchNormalization & (None, 63, 63, 31, 64) & 256\\
			\hline
			Conv3D & (None, 61, 61, 29, 64) & 110656\\
			\hline
 		MaxPooling3D & (None, 30, 30, 14, 64) & 0\\
			\hline
			BatchNormalization & (None, 30, 30, 14, 64) & 256\\
			\hline
			Conv3D & (None, 28, 28, 12, 128) & 221312\\
			\hline
			MaxPooling3D & (None, 14, 14, 6, 128) & 0\\
			\hline
			BatchNormalization & (None, 14, 14, 6, 128) & 512\\
			\hline
			Conv3D & (None, 12, 12, 4, 256) & 884992\\
			\hline
			MaxPooling3D & (None, 6, 6, 2, 256) & 0\\
			\hline
			BatchNormalization & (None, 6, 6, 2, 256) & 1024\\
			\hline
			GlobalMaxPooling3D & (None, 256) & 0\\
%			\hline
%			Dense & (None, 1024) & 132096\\
%			\hline
%			Dropout & (None, 1024) & 0\\
%			\hline
%			Dense & (None, 512) & 524800\\
%			\hline
%			Dropout & (None, 512) & 0\\
%			\hline
%			Dense & (None, 128) & 8256\\
%			\hline
%			Dropout & (None, 128) & 0\\
%			\hline
%			Dense & (None, 64) & 8256\\
%			\hline
%			Dense & (None, 4) & 260\\
%\hline
%			GlobalMaxPooling3D & (None, 256) & 0\\
		\hline
		\end{tabular}}
	\label{tab:parameters_CNN2}
\end{table}

\begin{figure*}[ht]
	\centering
	\includegraphics[width=1\textwidth]{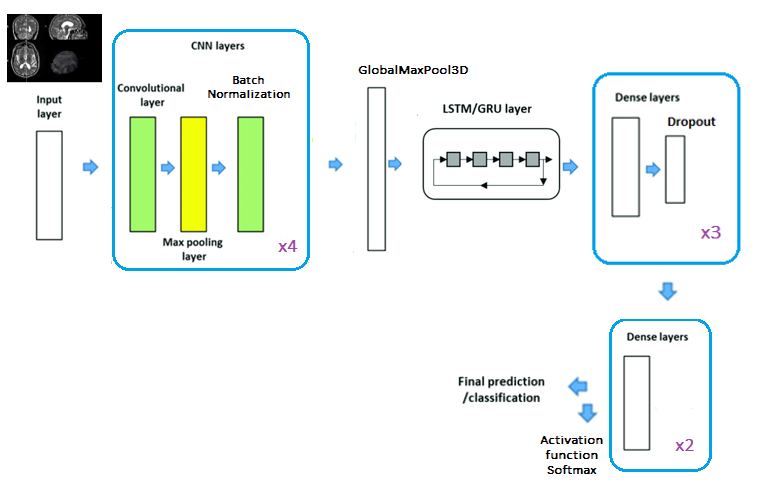}
	\caption{Neural network architecture without image merging.}
	\label{fig:reti2}
\end{figure*}

In particular, we adopt a network shown in Fig.~\ref{fig:reti2} that takes inspiration from 3DCNN + LSTM, widely used in many similar contexts, such as for the recognition of human gestures~\cite{zhang2017learning}~\cite{hakim2019dynamic} and to analyze the quality of a video~\cite{you2019deep}.
Summarizing, we can divide the network into three parts:

\begin{enumerate}
    \item In the first part we will find a 3DCNN that will be used to extract the characteristics of the three-dimensional MRI and will be contained in a Time Distributed layer that will allow us to implement the architecture described above. Details are shown in Table~\ref{tab:parameters_CNN2}.
    \item The second part of the network contains the layers to analyze the temporal information, the Time Distributed layer\footnote{\url{https://keras.io/api/layers/recurrent_layers/time_distributed/}}, which is a wrapper to apply a layer to every temporal slice of an input.
    \item The last part aims to provide the classification results.
\end{enumerate}

The 3DCNN+LSTM/GRU network has been trained and evaluated by dividing the dataset into a training set, and test set, respectively composed of 80\% and 20\% of the elements. 
We adopted k-fold validation, a resampling method that uses different portions of the data to test and train a model on different iterations. It is mainly used in contexts where the goal is to estimate how accurately a predictive model will work in practice\cite{valente2021cross}. The dataset is partitioned into k sets of equal size. Of the k sets, one is kept for testing the model~ and the rest are used as training data. The process is repeated k times, with each k set used exactly once for validation. Finally, their results are averaged to produce a single estimate~\cite{rafalo2022cross}.

We adopted k-fold validation with k=10.

The 3DCNN architecture is reported in Table~\ref{fig:reti2}. The last level produces a one-dimensional output to be used by the recurrent layers in the second part. We adopted a GlobalMaxPool3D layer for reducing the dimension getting the maximum values from the last convolution.
To this basic model of 3DCNN other characteristic layers of each network examined have been added, thus obtaining six different mixed neural networks. The added layers refer to:

\begin{itemize}
    \item \emph{Gated Recurrent Unit (GRU)}, shown in Table~\ref{tab:parameters_GRU}.
    
    \begin{table}[h!]
	\caption{GRU layer Parameters}
	\centering
	\resizebox{0.7\columnwidth}{!}{
		\begin{tabular}{|c|c|c|}
		\hline
			\textbf{Layer (type)} & \textbf{Output Shape}  & \textbf{Param} \\
			\hline
			GRU & [(None, 128)] & 148224 \\
			\hline
            Dense & [(None, 1024)] & 132096 \\
			\hline
            Dropout & [(None, 1024)] & 0 \\
			\hline
            Dense & [(None, 512)] & 524800 \\
			\hline
            Dropout & [(None, 512)] & 0 \\
			\hline
            Dense & [(None, 128)] & 65664 \\
			\hline
            Dropout & [(None, 128)] & 0 \\
			\hline
            Dense & [(None, 64)] & 8256 \\
			\hline
            Dense & [(None, 4)] & 260 \\
			\hline
		\end{tabular}}
	\label{tab:parameters_GRU}
    \end{table}

    \item \emph{sGRU}, composed by two stacked sGRU layers, as shown in Table~\ref{tab:parameters_sGRU}.
    
    \begin{table}[h!]
	\caption{sGRu layers Parameters}
	\centering
	\resizebox{0.7\columnwidth}{!}{
		\begin{tabular}{|c|c|c|}
		\hline
			\textbf{Layer (type)} & \textbf{Output Shape}  & \textbf{Param} \\
			\hline
			GRU & [(None, None, 128)] & 148224 \\
			\hline
			GRU & [(None, 128)] & 99072 \\
			\hline
            Dense & [(None, 1024)] & 132096 \\
			\hline
            Dropout & [(None, 1024)] & 0 \\
			\hline
            Dense & [(None, 512)] & 524800 \\
			\hline
            Dropout & [(None, 512)] & 0 \\
			\hline
            Dense & [(None, 128)] & 65664 \\
			\hline
            Dropout & [(None, 128)] & 0 \\
			\hline
            Dense & [(None, 64)] & 8256 \\
			\hline
            Dense & [(None, 4)] & 260 \\
			\hline
		\end{tabular}}
	\label{tab:parameters_sGRU}
    \end{table}
    
    \item\emph{sbiGRU}, composed of two bi-directional stacked GRU layers, as shown in Table~\ref{tab:parameters_sbiGRU}.
    
    \begin{table}[h!]
	\caption{sbiGRu layers Parameters}
	\centering
	\resizebox{0.7\columnwidth}{!}{
		\begin{tabular}{|c|c|c|}
		\hline
			\textbf{Layer (type)} & \textbf{Output Shape}  & \textbf{Param} \\
			\hline
			GRU & [(None, None, 256)] & 296448 \\
			\hline
			GRU & [(None, 256)] & 296448 \\
			\hline
            Dense & [(None, 1024)] & 263168 \\
			\hline
            Dropout & [(None, 1024)] & 0 \\
			\hline
            Dense & [(None, 512)] & 524800 \\
			\hline
            Dropout & [(None, 512)] & 0 \\
			\hline
            Dense & [(None, 128)] & 65664 \\
			\hline
            Dropout & [(None, 128)] & 0 \\
			\hline
            Dense & [(None, 64)] & 8256 \\
			\hline
            Dense & [(None, 4)] & 260 \\
			\hline
		\end{tabular}}
	\label{tab:parameters_sbiGRU}
    \end{table}
    
    \item\emph{LSTM}, shown in Table~\ref{tab:parameters_LSTM}. 
    
    \begin{table}[h!]
	\caption{LSTM layer Parameters}
	\centering
	\resizebox{0.7\columnwidth}{!}{
		\begin{tabular}{|c|c|c|}
		\hline
			\textbf{Layer (type)} & \textbf{Output Shape}  & \textbf{Param} \\
			\hline
			LSTM & [(None, 128)] & 197120 \\
			\hline
            Dense & [(None, 1024)] & 132096 \\
			\hline
            Dropout & [(None, 1024)] & 0 \\
			\hline
            Dense & [(None, 512)] & 524800 \\
			\hline
            Dropout & [(None, 512)] & 0 \\
			\hline
            Dense & [(None, 128)] & 65664 \\
			\hline
            Dropout & [(None, 128)] & 0 \\
			\hline
            Dense & [(None, 64)] & 8256 \\
			\hline
            Dense & [(None, 4)] & 260 \\
			\hline
		\end{tabular}}
	\label{tab:parameters_LSTM}
    \end{table}
    
    \item\emph{sLSTM}, composed of two stacked LSTM layers shown in Table~\ref{tab:parameters_sLSTM}. 
    
    \begin{table}[h!]
	\caption{sLSTM layers Parameters}
	\centering
	\resizebox{0.7\columnwidth}{!}{
		\begin{tabular}{|c|c|c|}
		\hline
			\textbf{Layer (type)} & \textbf{Output Shape}  & \textbf{Param} \\
			\hline
			LSTM & [(None, None, 128)] & 197120 \\
			\hline
			LSTM & [(None, 128)] & 131584 \\
			\hline
            Dense & [(None, 1024)] & 132096 \\
			\hline
            Dropout & [(None, 1024)] & 0 \\
			\hline
            Dense & [(None, 512)] & 524800 \\
			\hline
            Dropout & [(None, 512)] & 0 \\
			\hline
            Dense & [(None, 128)] & 65664 \\
			\hline
            Dropout & [(None, 128)] & 0 \\
			\hline
            Dense & [(None, 64)] & 8256 \\
			\hline
            Dense & [(None, 4)] & 260 \\
			\hline
		\end{tabular}}
	\label{tab:parameters_sLSTM}
    \end{table}
    
    \item\emph{sbiLSTM}: it consists in two LSTM layers stacked bidirectionally shown in Table~\ref{tab:parameters_sbiLSTM}.
    
    \begin{table}[h!]
	\caption{sbiLSTM layers Parameters}
	\centering
	\resizebox{0.7\columnwidth}{!}{
		\begin{tabular}{|c|c|c|}
		\hline
			\textbf{Layer (type)} & \textbf{Output Shape}  & \textbf{Param} \\
			\hline
			LSTM & [(None, None, 256)] & 394240 \\
			\hline
			LSTM & [(None, 256)] & 394240 \\
			\hline
            Dense & [(None, 1024)] & 263168 \\
			\hline
            Dropout & [(None, 1024)] & 0 \\
			\hline
            Dense & [(None, 512)] & 524800 \\
			\hline
            Dropout & [(None, 512)] & 0 \\
			\hline
            Dense & [(None, 128)] & 65664 \\
			\hline
            Dropout & [(None, 128)] & 0 \\
			\hline
            Dense & [(None, 64)] & 8256 \\
			\hline
            Dense & [(None, 4)] & 260 \\
			\hline
		\end{tabular}}
	\label{tab:parameters_sbiLSTM}
    \end{table}
    \end{itemize}

\subsubsection*{Performance metrics}
We aimed to compare the six considered Deep learning models to evaluate which configuration had the best classification performance in Parkinson's progression classes.

The ML performance metrics for classification tasks are generally based on the confusion matrix, an $NxN$ matrix used to evaluate the performance of a classification model, where N is the number of target classes. The matrix compares the actual values with those produced by the model, offering a holistic view of the performance of the classification model and the type of errors it is making. From the confusion matrices, it is possible to obtain the following values:

\begin{itemize}
    \item True Positives (TP): number of samples that the model correctly predicts in the positive class.
    \item True Negatives (TN): number of samples that the model correctly predicts in the negative class.
    \item False Positives (FP): number of samples that the model incorrectly predicts in the positive class.
    \item False Negatives (FN): number of samples that the model incorrectly predicts in the negative class.
\end{itemize}

\begin{figure*}[ht]
	\centering
	\includegraphics[width=0.9\textwidth]{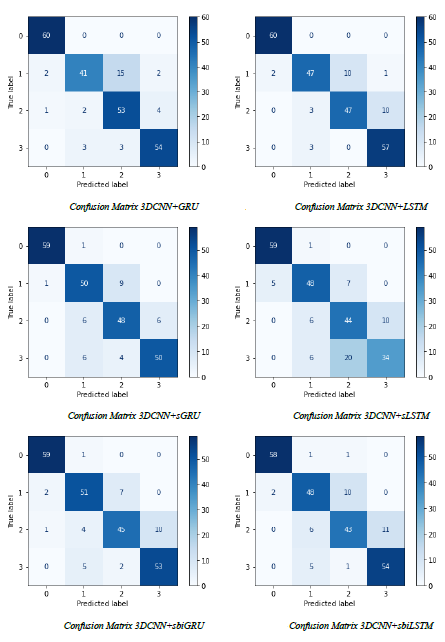}
	\caption{Confusion matrix.}
	\label{fig:confusion_matrix}
\end{figure*}

To assess the performance of the selected networks the multiclass confusion matrices are computed on the test set for each model. Fig.~\ref{fig:confusion_matrix} shows the multiclass confusion matrix computed by running the considered models.
For assessing the performance of the model the following metrics~\cite{berger2020threshold} based on the multiclass confusion matrix were computed. 

\begin{itemize}
    \item\emph{Macro Averaged Accuracy.} It is computed as the average of each accuracy per class.
    First, we calculate the accuracy for each class $k$ separately: 

\begin{equation}
Accuracy_{k}=
        \frac{TP_{k}}{TP_{k} +TN_{k}}
    \end{equation}
Then we compute the average of all accuracy values:

\begin{equation}
        MAAccuracy=\frac{\sum_{k=1}^{K} accuracy_{k}}{K}
    \end{equation}

    \item\emph{Macro Averaged Precision.} It is computed as follows.
    First, we calculate the precision for each class $k$ separately: 
    \begin{equation}
Precision_{k}=
        \frac{TP_{k}}{TP_{k} +FP_{k}}
    \end{equation}
    Then we compute the average of all precision values:

    \begin{equation}
       MAPrecision= \frac{\sum_{k=1}^{K} Precision_{k}}{K}
    \end{equation}

    \item\emph{Macro Averaged Recall.} It is computed as follows.
    First, we calculate the Recall for each class $k$ separately: 
    \begin{equation}
Recall_{k}=
        \frac{TP_{k}}{TP_{k} +FN_{k}}
    \end{equation}
    Then we compute the average of all recall values:
    \begin{equation}
       MARecall= \frac{\sum_{k=1}^{K} Recall_{k}}{K}
    \end{equation}

    \item\emph{Macro Averaged F1.} MAF1 is the weighted average (harmonic average) of precision and recall, computed as follows:
    %Therefore, this score takes into account both false positives and false negatives. It is calculated using the following formula:
    
    \begin{equation}
        2 * \left ( \frac{Macro Average Precision * Macro Average Recall}{Macro Average Precision^{-1} + Macro Average Recall^{-1}} \right )
    \end{equation}
    
    \item\emph{Macro Averaged OVR AUC.} It quantifies the model's ability to distinguish each class (One Versus Rest). To derive this metric we need to compute the AUC of each ROC curve for each class~\cite{kumar2022expression} and to compute the average.

\end{itemize}

 \begin{table*}[h!]
	\caption{Model performance comparison.}
	\centering
	\resizebox{1\columnwidth}{!}{
		\begin{tabular}{ |c|c|c|c|c|c|}
		\hline
			\textbf{Model} & \textbf{MAAccuracy Validation}  & \textbf{MAAccuracy Test}  & \textbf{MAPrecision}  & \textbf{MARecall}  & \textbf{MAF1}\\
			\hline
			3DCNN+GRU & 82\% & 86,6\% & 87,2\% & 86,6\% & 86,4\% \\
			\hline
			3DCNN+sGRU & 78\% & 86,2\% & 86,4\% & 86,2\% & 86,2\% \\
			\hline
			3DCNN+sbiGRU & 86\% & 86,6\% & 86,5\% & 86,6\% & 86,5\% \\
			\hline
			3DCNN+LSTM & 86\% & \textbf{87,9\%} & \textbf{87,9\%} & \textbf{87,9\%} & \textbf{87,7\%} \\
			\hline
			3DCNN+sLSTM & 70\% & 77\% & 77,5\% & 77\% & 76,7\% \\
			\hline
			3DCNN+sbiLSTM & 81\% & 84,5\% & 84,4\% & 84,5\% & 84.4\% \\
			\hline
		\end{tabular}}
	\label{tab:Results}
    \end{table*}

 \begin{table}[h!]
	\caption{Results Macro Averaged OVR AUC.}
	\centering
	\resizebox{1\columnwidth}{!}{
		\begin{tabular}{|c|c|c|c|c|c|}
		\hline
			\textbf{Experiment} & \textbf{0}  & \textbf{1}  & \textbf{2}  & \textbf{3}& \textbf{MA OVR AUC}\\
			\hline
			3DCNN+GRU & 99,1\% & 82,7\% & 89,1\% & 93,3\%& 91.05\%\\
			\hline
			3DCNN+sGRU & 98,8\% & 88\% & 86,3\% & 90\%&90.76\%\\ 
			\hline
			3DCNN+sbiGRU & 98,3\% & 89,7\% & 85\% & 91,3\%&91.08\%\\
			\hline
			3DCNN+LSTM & 99,4\% & 87,5\% & 86,3\% & 94,4\%&\textbf{91.90\%}\\ 
			\hline
			3DCNN+sLSTM & 97,5\% & 86,3\% & 79,1\% & 75,50\%&84.60\%\\
			\hline
			3DCNN+sbiLSTM & 97,7\% & 86,6\% & 82,5\% & 91,90\%&89.68\%\\ 
			\hline
		\end{tabular}}
	\label{tab:Results2}
    \end{table}

\section{Models' assessment}
\label{results}

Results of the performance of the six models on the test set are detailed in the following and summarized in Tables~\ref{tab:Results} and~\ref{tab:Results2}. All the models have been trained for 35 epochs.
   
As shown in Table~\ref{tab:Results} on the test set the 3DCNN + LSTM network performs better than the others, even if the results are close, except for 3DCNN+sLSTM which has the lowest performance.  From the AUC values for each class (Tab.~\ref{tab:Results2}), it can be seen that all the models can recognize class 0 very well, compared to class 2 where there is more variance between the various results. The best Macro Averaged OVR AUC result, 91.9\%, is still reached by 3DCNN+LSTM, whose performance varies from  86.3\% in the case of the class 2 to 99.4\% for the class 0.

\section*{Conclusion and Future Work}
\label{sec:conclusion}
Our research centred on developing a deep neural network-based novel model to detect and assess Parkinson's disease progression. We created and trained a model that combines a recurrent layer (LSTM or GRU) with a 3D convolutional neural network (3DCNN) using analysis of the Parkinson's Progression Markers Initiative dataset.

The outcomes show how good our method is; specifically, on four classes of Parkinson's disease development, the 3DCNN + LSTM model acquired a Macro Averaged OVR AUC of 91.9\%. This performance outperforms other tested models, highlighting the validity of our architecture in accurately classifying different stages of progression.

The precision of the model was enhanced by careful consideration given to data preprocessing, which included removing non-brain content and fixing acquisition mistakes. To maintain consistency in the results throughout the various advancement classes, the dataset was balanced by the application of template-model-based data augmentation approaches.

Nonetheless, we acknowledge that our research has certain limitations, such as reliance on the dataset's data and the requirement for additional model validation using a wide range of clinical data. To increase model accuracy, future advancements might investigate other deep learning strategies and use multimodal data.

In summary, our study constitutes a major advancement in the field of Parkinson's disease diagnosis and assessment. Adopting cutting-edge techniques based on deep neural networks provides a strong foundation for future advancements and therapeutic applications while opening up new avenues for understanding and controlling the illness

\nolinenumbers

%This is where your bibliography is generated. Make sure that your .bib file is actually called library.bib
\bibliography{library}

\begin{thebibliography}{10}

\bibitem{adeli2016joint}
E.~Adeli, F.~Shi, L.~An, C.-Y. Wee, G.~Wu, T.~Wang, and D.~Shen.
\newblock Joint feature-sample selection and robust diagnosis of parkinson's disease from mri data.
\newblock {\em NeuroImage}, 141:206--219, 2016.

\bibitem{aderghal2017classification}
K.~Aderghal, M.~Boissenin, J.~Benois-Pineau, G.~Catheline, and K.~Afdel.
\newblock Classification of smri for ad diagnosis with convolutional neuronal networks: A pilot 2-d+ $\backslash$epsilon$ $study on adni.
\newblock In {\em International Conference on Multimedia Modeling}, pages 690--701. Springer, 2017.

\bibitem{badvza2020classification}
M.~M. Bad{\v{z}}a and M.~{\v{C}}. Barjaktarovi{\'c}.
\newblock Classification of brain tumors from mri images using a convolutional neural network.
\newblock {\em Applied Sciences}, 10(6):1999, 2020.

\bibitem{bakas2018identifying}
S.~Bakas, M.~Reyes, A.~Jakab, S.~Bauer, M.~Rempfler, A.~Crimi, R.~T. Shinohara, C.~Berger, S.~M. Ha, M.~Rozycki, et~al.
\newblock Identifying the best machine learning algorithms for brain tumor segmentation, progression assessment, and overall survival prediction in the brats challenge.
\newblock {\em arXiv preprint arXiv:1811.02629}, 2018.

\bibitem{balestrino2020parkinson}
R.~Balestrino and A.~Schapira.
\newblock Parkinson disease.
\newblock {\em European journal of neurology}, 27(1):27--42, 2020.

\bibitem{berger2020threshold}
A.~Berger and S.~Guda.
\newblock Threshold optimization for f measure of macro-averaged precision and recall.
\newblock {\em Pattern Recognition}, 102:107250, 2020.

\bibitem{bhidayasiri2012parkinson}
R.~Bhidayasiri, D.~Tarsy, R.~Bhidayasiri, and D.~Tarsy.
\newblock Parkinson’s disease: Hoehn and yahr scale.
\newblock {\em Movement disorders: a video atlas: a video atlas}, pages 4--5, 2012.

\bibitem{bhurane2022diagnosis}
A.~A. Bhurane, S.~Dhok, M.~Sharma, R.~Yuvaraj, M.~Murugappan, and U.~R. Acharya.
\newblock Diagnosis of parkinson's disease from electroencephalography signals using linear and self-similarity features.
\newblock {\em Expert Systems}, 39(7):e12472, 2022.

\bibitem{bolas2010basic}
N.~Bolas.
\newblock Basic mri principles.
\newblock {\em Equine MRI}, pages 1--37, 2010.

\bibitem{braak2000pathoanatomy}
H.~Braak and E.~Braak.
\newblock Pathoanatomy of parkinson’s disease.
\newblock {\em Journal of neurology}, 247:II3--II10, 2000.

\bibitem{bui2022virtual}
V.~Bui and A.~Alaei.
\newblock Virtual reality in training artificial intelligence-based systems: a case study of fall detection.
\newblock {\em Multimedia Tools and Applications}, pages 1--18, 2022.

\bibitem{burciu2017progression}
R.~G. Burciu, E.~Ofori, D.~B. Archer, S.~S. Wu, O.~Pasternak, N.~R. McFarland, M.~S. Okun, and D.~E. Vaillancourt.
\newblock Progression marker of parkinson’s disease: a 4-year multi-site imaging study.
\newblock {\em Brain}, 140(8):2183--2192, 2017.

\bibitem{butt2018objective}
A.~H. Butt, E.~Rovini, C.~Dolciotti, G.~De~Petris, P.~Bongioanni, M.~Carboncini, and F.~Cavallo.
\newblock Objective and automatic classification of parkinson disease with leap motion controller.
\newblock {\em Biomedical engineering online}, 17(1):1--21, 2018.

\bibitem{can2020gating}
T.~Can, K.~Krishnamurthy, and D.~J. Schwab.
\newblock Gating creates slow modes and controls phase-space complexity in grus and lstms.
\newblock In {\em Mathematical and Scientific Machine Learning}, pages 476--511. PMLR, 2020.

\bibitem{chlap2021review}
P.~Chlap, H.~Min, N.~Vandenberg, J.~Dowling, L.~Holloway, and A.~Haworth.
\newblock A review of medical image data augmentation techniques for deep learning applications.
\newblock {\em Journal of Medical Imaging and Radiation Oncology}, 65(5):545--563, 2021.

\bibitem{chu2002age}
Y.~Chu, K.~Kompoliti, E.~J. Cochran, E.~J. Mufson, and J.~H. Kordower.
\newblock Age-related decreases in nurr1 immunoreactivity in the human substantia nigra.
\newblock {\em Journal of Comparative Neurology}, 450(3):203--214, 2002.

\bibitem{cleary2015deep}
D.~R. Cleary, A.~Ozpinar, A.~M. Raslan, and A.~L. Ko.
\newblock Deep brain stimulation for psychiatric disorders: where we are now.
\newblock {\em Neurosurgical focus}, 38(6):E2, 2015.

\bibitem{dadu2022identification}
A.~Dadu, V.~Satone, R.~Kaur, S.~H. Hashemi, H.~Leonard, H.~Iwaki, M.~B. Makarious, K.~J. Billingsley, S.~Bandres-Ciga, L.~J. Sargent, et~al.
\newblock Identification and prediction of parkinson’s disease subtypes and progression using machine learning in two cohorts.
\newblock {\em npj Parkinson's Disease}, 8(1):172, 2022.

\bibitem{dakof1986parkinson}
G.~A. Dakof and G.~A. Mendelsohn.
\newblock Parkinson's disease: the psychological aspects of a chronic illness.
\newblock {\em Psychological Bulletin}, 99(3):375, 1986.

\bibitem{dubey2011evaluation}
R.~Dubey, M.~Hanmandlu, and S.~Vasikarla.
\newblock Evaluation of three methods for mri brain tumor segmentation.
\newblock In {\em 2011 eighth international conference on information technology: new generations}, pages 494--499. IEEE, 2011.

\bibitem{ferguson2016early}
L.~W. Ferguson, A.~H. Rajput, and A.~Rajput.
\newblock Early-onset vs. late-onset parkinson’s disease: A clinical-pathological study.
\newblock {\em Canadian journal of neurological sciences}, 43(1):113--119, 2016.

\bibitem{ferreira2022machine}
M.~I.~A. Ferreira, F.~A. Barbieri, V.~C. Moreno, T.~Penedo, and J.~M.~R. Tavares.
\newblock Machine learning models for parkinson’s disease detection and stage classification based on spatial-temporal gait parameters.
\newblock {\em Gait \& Posture}, 98:49--55, 2022.

\bibitem{forbes2021processing}
S.~H. Forbes, S.~Wijeakumar, A.~T. Eggebrecht, V.~A. Magnotta, and J.~P. Spencer.
\newblock Processing pipeline for image reconstructed fnirs analysis using both mri templates and individual anatomy.
\newblock {\em Neurophotonics}, 8(2):025010, 2021.

\bibitem{frasca2022visualizing}
M.~Frasca and G.~Tortora.
\newblock Visualizing correlations among parkinson biomedical data through information retrieval and machine learning techniques.
\newblock {\em Multimedia Tools and Applications}, 81(11):14685--14703, 2022.

\bibitem{gao2016deep}
Y.~Gao and D.~Glowacka.
\newblock Deep gate recurrent neural network.
\newblock In {\em Asian conference on machine learning}, pages 350--365. PMLR, 2016.

\bibitem{garcia2023determining}
M.~T. Garc{\'\i}a-Ord{\'a}s, J.~A. Ben{\'\i}tez-Andrades, J.~Aveleira-Mata, J.-M. Alija-P{\'e}rez, and C.~Benavides.
\newblock Determining the severity of parkinson’s disease in patients using a multi task neural network.
\newblock {\em Multimedia Tools and Applications}, pages 1--16, 2023.

\bibitem{gorgolewski2011nipype}
K.~Gorgolewski, C.~D. Burns, C.~Madison, D.~Clark, Y.~O. Halchenko, M.~L. Waskom, and S.~S. Ghosh.
\newblock Nipype: a flexible, lightweight and extensible neuroimaging data processing framework in python.
\newblock {\em Frontiers in neuroinformatics}, page~13, 2011.

\bibitem{greve2009accurate}
D.~N. Greve and B.~Fischl.
\newblock Accurate and robust brain image alignment using boundary-based registration.
\newblock {\em Neuroimage}, 48(1):63--72, 2009.

\bibitem{hakim2019dynamic}
N.~L. Hakim, T.~K. Shih, S.~P. Kasthuri~Arachchi, W.~Aditya, Y.-C. Chen, and C.-Y. Lin.
\newblock Dynamic hand gesture recognition using 3dcnn and lstm with fsm context-aware model.
\newblock {\em Sensors}, 19(24):5429, 2019.

\bibitem{hao2021comprehensive}
R.~Hao, K.~Namdar, L.~Liu, M.~A. Haider, and F.~Khalvati.
\newblock A comprehensive study of data augmentation strategies for prostate cancer detection in diffusion-weighted mri using convolutional neural networks.
\newblock {\em Journal of Digital Imaging}, 34(4):862--876, 2021.

\bibitem{hawkes2010timeline}
C.~H. Hawkes, K.~Del~Tredici, and H.~Braak.
\newblock A timeline for parkinson's disease.
\newblock {\em Parkinsonism \& related disorders}, 16(2):79--84, 2010.

\bibitem{he2021db}
J.-Y. He, X.~Wu, Z.-Q. Cheng, Z.~Yuan, and Y.-G. Jiang.
\newblock Db-lstm: Densely-connected bi-directional lstm for human action recognition.
\newblock {\em Neurocomputing}, 444:319--331, 2021.

\bibitem{heim2017magnetic}
B.~Heim, F.~Krismer, R.~De~Marzi, and K.~Seppi.
\newblock Magnetic resonance imaging for the diagnosis of parkinson’s disease.
\newblock {\em Journal of neural transmission}, 124(8):915--964, 2017.

\bibitem{hoehn1967parkinsonism}
M.~M. Hoehn and M.~D. Yahr.
\newblock Parkinsonism: onset, progression, and mortality.
\newblock {\em Neurology}, 17(5):427--427, 1967.

\bibitem{igwe2022automatic}
K.~C. Igwe, P.~J. Lao, R.~S. Vorburger, A.~Banerjee, A.~Rivera, A.~Chesebro, K.~Laing, J.~J. Manly, and A.~M. Brickman.
\newblock Automatic quantification of white matter hyperintensities on t2-weighted fluid attenuated inversion recovery magnetic resonance imaging.
\newblock {\em Magnetic Resonance Imaging}, 85:71--79, 2022.

\bibitem{inthavong2021list}
K.~Inthavong et~al.
\newblock List of useful computational software.
\newblock {\em Clinical and Biomedical Engineering in the Human Nose}, page 301, 2021.

\bibitem{jones2010twenty}
D.~K. Jones and M.~Cercignani.
\newblock Twenty-five pitfalls in the analysis of diffusion mri data.
\newblock {\em NMR in Biomedicine}, 23(7):803--820, 2010.

\bibitem{kharb2021review}
A.~Kharb and P.~Chaudhary.
\newblock A review on skull stripping techniques of brain mri images.
\newblock {\em Webology (ISSN: 1735-188X)}, 18(6), 2021.

\bibitem{khaskhoussy2023improving}
R.~Khaskhoussy and Y.~B. Ayed.
\newblock Improving parkinson’s disease recognition through voice analysis using deep learning.
\newblock {\em Pattern Recognition Letters}, 2023.

\bibitem{kinney2019high}
B.~M. Kinney and P.~Lozanova.
\newblock High intensity focused electromagnetic therapy evaluated by magnetic resonance imaging: Safety and efficacy study of a dual tissue effect based non-invasive abdominal body shaping.
\newblock {\em Lasers in surgery and medicine}, 51(1):40--46, 2019.

\bibitem{kouli2018parkinson}
A.~Kouli, K.~M. Torsney, and W.-L. Kuan.
\newblock Parkinson’s disease: etiology, neuropathology, and pathogenesis.
\newblock {\em Exon Publications}, pages 3--26, 2018.

\bibitem{kriegeskorte2019neural}
N.~Kriegeskorte and T.~Golan.
\newblock Neural network models and deep learning.
\newblock {\em Current Biology}, 29(7):R231--R236, 2019.

\bibitem{kumar2022expression}
P.~Kumar, L.~A. Raj, K.~M. Sagayam, and N.~S. Ram.
\newblock Expression invariant face recognition based on multi-level feature fusion and transfer learning technique.
\newblock {\em Multimedia Tools and Applications}, pages 1--19, 2022.

\bibitem{li2021moving}
X.~Li, L.~Ai, S.~Giavasis, H.~Jin, E.~Feczko, T.~Xu, J.~Clucas, A.~Franco, A.~S. Heinsfeld, A.~Adebimpe, et~al.
\newblock Moving beyond processing and analysis-related variation in neuroscience.
\newblock {\em BioRxiv}, 2021.

\bibitem{lu20193d}
H.~Lu, H.~Wang, Q.~Zhang, S.~W. Yoon, and D.~Won.
\newblock A 3d convolutional neural network for volumetric image semantic segmentation.
\newblock {\em Procedia Manufacturing}, 39:422--428, 2019.

\bibitem{maiti2017current}
P.~Maiti, J.~Manna, and G.~L. Dunbar.
\newblock Current understanding of the molecular mechanisms in parkinson's disease: Targets for potential treatments.
\newblock {\em Translational neurodegeneration}, 6(1):1--35, 2017.

\bibitem{mani2013survey}
V.~Mani and S.~Arivazhagan.
\newblock Survey of medical image registration.
\newblock {\em Journal of Biomedical Engineering and Technology}, 1(2):8--25, 2013.

\bibitem{marek2011parkinson}
K.~Marek, D.~Jennings, S.~Lasch, A.~Siderowf, C.~Tanner, T.~Simuni, C.~Coffey, K.~Kieburtz, E.~Flagg, S.~Chowdhury, et~al.
\newblock The parkinson progression marker initiative (ppmi).
\newblock {\em Progress in neurobiology}, 95(4):629--635, 2011.

\bibitem{marwa2022mri}
E.-G. Marwa, H.~E.-D. Moustafa, F.~Khalifa, H.~Khater, and E.~AbdElhalim.
\newblock An mri-based deep learning approach for accurate detection of alzheimer’s disease.
\newblock {\em Alexandria Engineering Journal}, 2022.

\bibitem{modi2021deep}
H.~Modi, J.~Hathaliya, M.~S. Obaidiat, R.~Gupta, and S.~Tanwar.
\newblock Deep learning-based parkinson disease classification using pet scan imaging data.
\newblock In {\em 2021 IEEE 6th International Conference on Computing, Communication and Automation (ICCCA)}, pages 837--841. IEEE, 2021.

\bibitem{nalepa2019data}
J.~Nalepa, G.~Mrukwa, S.~Piechaczek, P.~R. Lorenzo, M.~Marcinkiewicz, B.~Bobek-Billewicz, P.~Wawrzyniak, P.~Ulrych, J.~Szymanek, M.~Cwiek, et~al.
\newblock Data augmentation via image registration.
\newblock In {\em 2019 IEEE international conference on image processing (ICIP)}, pages 4250--4254. IEEE, 2019.

\bibitem{nutt2005diagnosis}
J.~G. Nutt and G.~F. Wooten.
\newblock Diagnosis and initial management of parkinson's disease.
\newblock {\em New England Journal of Medicine}, 353(10):1021--1027, 2005.

\bibitem{pei2022general}
L.~Pei, M.~Ak, N.~H.~M. Tahon, S.~Zenkin, S.~Alkarawi, A.~Kamal, M.~Yilmaz, L.~Chen, M.~Er, N.~Ak, et~al.
\newblock A general skull stripping of multiparametric brain mris using 3d convolutional neural network.
\newblock {\em Scientific Reports}, 12(1):1--11, 2022.

\bibitem{peng2017multilevel}
B.~Peng, S.~Wang, Z.~Zhou, Y.~Liu, B.~Tong, T.~Zhang, and Y.~Dai.
\newblock A multilevel-roi-features-based machine learning method for detection of morphometric biomarkers in parkinson’s disease.
\newblock {\em Neuroscience letters}, 651:88--94, 2017.

\bibitem{poewe2008non}
W.~Poewe.
\newblock Non-motor symptoms in parkinson’s disease.
\newblock {\em European journal of neurology}, 15:14--20, 2008.

\bibitem{prashanth2016high}
R.~Prashanth, S.~D. Roy, P.~K. Mandal, and S.~Ghosh.
\newblock High-accuracy classification of parkinson's disease through shape analysis and surface fitting in 123i-ioflupane spect imaging.
\newblock {\em IEEE journal of biomedical and health informatics}, 21(3):794--802, 2016.

\bibitem{priyadarshi2001environmental}
A.~Priyadarshi, S.~A. Khuder, E.~A. Schaub, and S.~S. Priyadarshi.
\newblock Environmental risk factors and parkinson's disease: a metaanalysis.
\newblock {\em Environmental research}, 86(2):122--127, 2001.

\bibitem{rafalo2022cross}
M.~Rafa{\l}o.
\newblock Cross validation methods: analysis based on diagnostics of thyroid cancer metastasis.
\newblock {\em ICT Express}, 8(2):183--188, 2022.

\bibitem{rajanbabu2022ensemble}
K.~Rajanbabu, I.~K. Veetil, V.~Sowmya, E.~Gopalakrishnan, and K.~Soman.
\newblock Ensemble of deep transfer learning models for parkinson's disease classification.
\newblock In {\em Soft Computing and Signal Processing}, pages 135--143. Springer, 2022.

\bibitem{ramasamy2021segmentation}
J.~Ramasamy, R.~Doshi, and K.~K. Hiran.
\newblock Segmentation of brain tumor using deep learning methods: A review.
\newblock In {\em Proceedings of the International Conference on Data Science, Machine Learning and Artificial Intelligence}, pages 209--215, 2021.

\bibitem{rejusha2021artificial}
T.~Rejusha and V.~K. KS.
\newblock Artificial mri image generation using deep convolutional gan and its comparison with other augmentation methods.
\newblock In {\em 2021 international conference on communication, Control and Information Sciences (ICCISc)}, volume~1, pages 1--6. IEEE, 2021.

\bibitem{rewar2015systematic}
S.~Rewar.
\newblock A systematic review on parkinson's disease (pd).
\newblock {\em Indian Journal of Research in Pharmacy and Biotechnology}, 3(2):176, 2015.

\bibitem{rosenblum2013handwriting}
S.~Rosenblum, M.~Samuel, S.~Zlotnik, I.~Erikh, and I.~Schlesinger.
\newblock Handwriting as an objective tool for parkinson’s disease diagnosis.
\newblock {\em Journal of neurology}, 260:2357--2361, 2013.

\bibitem{safdar2020comparative}
M.~F. Safdar, S.~S. Alkobaisi, and F.~T. Zahra.
\newblock A comparative analysis of data augmentation approaches for magnetic resonance imaging (mri) scan images of brain tumor.
\newblock {\em Acta informatica medica}, 28(1):29, 2020.

\bibitem{sajjad2019multi}
M.~Sajjad, S.~Khan, K.~Muhammad, W.~Wu, A.~Ullah, and S.~W. Baik.
\newblock Multi-grade brain tumor classification using deep cnn with extensive data augmentation.
\newblock {\em Journal of computational science}, 30:174--182, 2019.

\bibitem{saravanan2022systematic}
S.~Saravanan, K.~Ramkumar, K.~Adalarasu, V.~Sivanandam, S.~R. Kumar, S.~Stalin, and R.~Amirtharajan.
\newblock A systematic review of artificial intelligence (ai) based approaches for the diagnosis of parkinson’s disease.
\newblock {\em Archives of Computational Methods in Engineering}, 29(6):3639--3653, 2022.

\bibitem{scatton1982dopamine}
B.~Scatton, L.~Rouquier, F.~Javoy-Agid, and Y.~Agid.
\newblock Dopamine deficiency in the cerebral cortex in parkinson disease.
\newblock {\em Neurology}, 32(9):1039--1039, 1982.

\bibitem{severson2021discovery}
K.~A. Severson, L.~M. Chahine, L.~A. Smolensky, M.~Dhuliawala, M.~Frasier, K.~Ng, S.~Ghosh, and J.~Hu.
\newblock Discovery of parkinson's disease states and disease progression modelling: a longitudinal data study using machine learning.
\newblock {\em The Lancet Digital Health}, 3(9):e555--e564, 2021.

\bibitem{shahid2020deep}
A.~H. Shahid and M.~P. Singh.
\newblock A deep learning approach for prediction of parkinson’s disease progression.
\newblock {\em Biomedical Engineering Letters}, 10(2):227--239, 2020.

\bibitem{sharma2020automated}
H.~Sharma, S.~Soltaninejad, and I.~Cheng.
\newblock Automated classification of parkinson’s disease using diffusion tensor imaging data.
\newblock In {\em International Symposium on Visual Computing}, pages 658--669. Springer, 2020.

\bibitem{shewalkar2019performance}
A.~Shewalkar.
\newblock Performance evaluation of deep neural networks applied to speech recognition: Rnn, lstm and gru.
\newblock {\em Journal of Artificial Intelligence and Soft Computing Research}, 9(4):235--245, 2019.

\bibitem{singh20203d}
S.~P. Singh, L.~Wang, S.~Gupta, H.~Goli, P.~Padmanabhan, and B.~Guly{\'a}s.
\newblock 3d deep learning on medical images: a review.
\newblock {\em Sensors}, 20(18):5097, 2020.

\bibitem{smith2000bet}
S.~M. Smith.
\newblock Bet: Brain extraction tool.
\newblock {\em FMRIB TR00SMS2b, Oxford Centre for Functional Magnetic Resonance Imaging of the Brain), Department of Clinical Neurology, Oxford University, John Radcliffe Hospital, Headington, UK}, 2000.

\bibitem{strother2006evaluating}
S.~C. Strother.
\newblock Evaluating fmri preprocessing pipelines.
\newblock {\em IEEE Engineering in Medicine and Biology Magazine}, 25(2):27--41, 2006.

\bibitem{templeton2022classification}
J.~M. Templeton, C.~Poellabauer, and S.~Schneider.
\newblock Classification of parkinson’s disease and its stages using machine learning.
\newblock {\em Scientific Reports}, 12(1):14036, 2022.

\bibitem{truong2011bradley}
D.~D. Truong, W.~M. Carroll, and R.~Bhidayasiri.
\newblock Bradley j. robottom, william j. weiner, and lisa m. shulman.
\newblock {\em International Neurology: A Clinical Approach}, page 152, 2011.

\bibitem{valente2021cross}
G.~Valente, A.~L. Castellanos, L.~Hausfeld, F.~De~Martino, and E.~Formisano.
\newblock Cross-validation and permutations in mvpa: Validity of permutation strategies and power of cross-validation schemes.
\newblock {\em NeuroImage}, 238:118145, 2021.

\bibitem{veetil2021parkinson}
I.~K. Veetil, E.~Gopalakrishnan, V.~Sowmya, and K.~Soman.
\newblock Parkinson’s disease classification from magnetic resonance images (mri) using deep transfer learned convolutional neural networks.
\newblock In {\em 2021 IEEE 18th India Council International Conference (INDICON)}, pages 1--6. IEEE, 2021.

\bibitem{vyas2022deep}
T.~Vyas, R.~Yadav, C.~Solanki, R.~Darji, S.~Desai, and S.~Tanwar.
\newblock Deep learning-based scheme to diagnose parkinson's disease.
\newblock {\em Expert Systems}, 39(3):e12739, 2022.

\bibitem{wahid2015classification}
F.~Wahid, R.~K. Begg, C.~J. Hass, S.~Halgamuge, and D.~C. Ackland.
\newblock Classification of parkinson's disease gait using spatial-temporal gait features.
\newblock {\em IEEE journal of biomedical and health informatics}, 19(6):1794--1802, 2015.

\bibitem{wakabayashi2020and}
K.~Wakabayashi.
\newblock Where and how alpha-synuclein pathology spreads in parkinson’s disease.
\newblock {\em Neuropathology}, 40(5):415--425, 2020.

\bibitem{wodzinski2019deep}
M.~Wodzinski, A.~Skalski, D.~Hemmerling, J.~R. Orozco-Arroyave, and E.~N{\"o}th.
\newblock Deep learning approach to parkinson’s disease detection using voice recordings and convolutional neural network dedicated to image classification.
\newblock In {\em 2019 41st Annual International Conference of the IEEE Engineering in Medicine and Biology Society (EMBC)}, pages 717--720. IEEE, 2019.

\bibitem{you2019deep}
J.~You and J.~Korhonen.
\newblock Deep neural networks for no-reference video quality assessment.
\newblock In {\em 2019 IEEE International Conference on Image Processing (ICIP)}, pages 2349--2353. IEEE, 2019.

\bibitem{zeng2017differentiating}
L.-L. Zeng, L.~Xie, H.~Shen, Z.~Luo, P.~Fang, Y.~Hou, B.~Tang, T.~Wu, and D.~Hu.
\newblock Differentiating patients with parkinson’s disease from normal controls using gray matter in the cerebellum.
\newblock {\em The Cerebellum}, 16(1):151--157, 2017.

\bibitem{zhang2019explainable}
A.~Y. Zhang, S.~S.~W. Lam, M.~E.~H. Ong, P.~H. Tang, and L.~L. Chan.
\newblock Explainable ai: classification of mri brain scans orders for quality improvement.
\newblock In {\em Proceedings of the 6th IEEE/ACM International Conference on Big Data Computing, Applications and Technologies}, pages 95--102, 2019.

\bibitem{zhang2017learning}
L.~Zhang, G.~Zhu, P.~Shen, J.~Song, S.~Afaq~Shah, and M.~Bennamoun.
\newblock Learning spatiotemporal features using 3dcnn and convolutional lstm for gesture recognition.
\newblock In {\em Proceedings of the IEEE International Conference on Computer Vision Workshops}, pages 3120--3128, 2017.

\end{thebibliography}

%This defines the bibliographies style. Search online for a list of available styles.
\bibliographystyle{abbrv}

\end{document}